\documentclass[iop]{emulateapj}
\usepackage{apjfonts}

\begin{document}

\title{Analysis of GRB 080319B and GRB 050904 within the fireshell model: evidence for a broader spectral energy distribution}

\author{
B. Patricelli\altaffilmark{1,2,3}, M.G. Bernardini\altaffilmark{1,2,4}, C.L. Bianco\altaffilmark{1,2}, L. Caito\altaffilmark{1,2}, G. de Barros\altaffilmark{1,2}, L. Izzo\altaffilmark{1,2}, R. Ruffini\altaffilmark{1,2,5}, G.V. Vereshchagin\altaffilmark{1,2}
}

\altaffiltext{1}{ICRANet, Piazzale della Repubblica 10, I-65122 Pescara, Italy. E-mails: [barbara.patricelli;gustavo.debarros]@icranet.org.}
\altaffiltext{2}{ICRA and Dipartimento di Fisica, Universit\`a di Roma ``La Sapienza'', Piazzale Aldo Moro 5, I-00185 Roma, Italy. E-mails: [maria.bernardini;bianco;letizia.caito;luca.izzo;ruffini;veresh]@icra.it.}
\altaffiltext{3}{Astronomy Institute - UNAM, Ciudad Universitaria C.P. 04510 M\'exico D.F., M\'exico.}
\altaffiltext{4}{INAF - Osservatorio Astronomico di Brera, via Emilio Bianchi 46, 23807 Merate, Italy.}
\altaffiltext{5}{ICRANet, Universit\'e de Nice Sophia Antipolis, Grand Ch\^ateau, BP 2135, 28, avenue de Valrose, 06103 NICE CEDEX 2, France.}

\shorttitle{Analysis of GRB 080319B and GRB 050904 within the fireshell model: evidence for a broader SED}

\shortauthors{Patricelli et al.}

\begin{abstract}
The observation of GRB 080319B, with an isotropic energy $E_{iso}=1.32 \times 10^{54}$ erg, and GRB 050904, with $E_{iso}=1.04 \times 10^{54}$ erg, offers the possibility of studying the spectral properties of the prompt radiation of two of the most energetic Gamma Ray Bursts (GRBs). This allows us to probe the validity of the fireshell model for GRBs beyond $10^{54}$ erg, well outside the energy range where it has been successfully tested up to now ($10^{49} $--$ 10^{53}$ erg). We find that in the low energy region, the prompt emission spectra observed by \emph{Swift} BAT reveals more power than theoretically predicted. The opportunities offered by these observations to improve the fireshell model are outlined in this paper. One of the distinguishing features of the fireshell model is that it relates the observed GRB spectra to the spectrum in the comoving frame of the fireshell. Originally, a fully radiative condition and a comoving thermal spectrum were adopted. An additional power-law in the comoving thermal spectrum is required due to the discrepancy of the theoretical and observed light curves and spectra in the fireshell model for GRBs 080319B and 050904. A new phenomenological parameter $\alpha$ is correspondingly introduced in the model. We perform numerical simulations of the prompt emission in the \emph{Swift} BAT bandpass by assuming different values of $\alpha$ within the fireshell model. We compare them with the GRB 080319B and GRB 050904 observed time-resolved spectra, as well as with their time-integrated spectra and light curves. Although GRB 080319B and GRB 050904 are at very different redshifts (z=0.937 and z=6.29 respectively), a value of $\alpha=-1.8$ leads for both of them to a good agreement between the numerical simulations and the observed BAT light curves, time-resolved and time-integrated spectra. Such a modified spectrum is also consistent with the observations of previously analyzed less energetic GRBs and reasons for this additional agreement are given. Perspectives for future low energy missions are outlined.
\end{abstract}

\keywords{Gamma-ray burst: general --- Gamma-ray burst: individual: GRB 080319B --- Gamma-ray burst: individual: GRB 050904 --- ISM: structure --- Black hole physics}

\section{Introduction}\label{intro}

Out of the hundreds of Gamma Ray Bursts (GRBs) so far observed with known redshifts, there are approximately ten GRBs with an isotropic energy $E_{iso} \gtrsim 10^{54}$ erg: GRB 990123, GRB 990506, GRB 000131, GRB 050820A, GRB 050904, GRB 080319B, GRB 080607, GRB 080721, GRB 080916C, GRB 090323, GRB 090926A \citep[see e.g.][]{2008MNRAS.391..577A,2009A&A...508..173A,2010ApJ...720.1513K}, GRB 090902B \citep[see][]{2009ApJ...706L.138A} and GRB 110918A \citep{2011GCN..12370...1F}. We will analyse two of these sources to probe the fireshell model, which has been successfully applied to GRBs with $E_{iso}$ up to $10^{53}$ erg. The two candidates are GRB 080319B and GRB 050904, having an isotropic $\gamma$-ray energy release respectively of $E_{iso}=1.32\times 10^{54}$ erg (20 keV -- 7 MeV, see \citealp{2008GCN..7482....1G}) and $E_{iso}=1.04^{+0.25}_{-0.17}\times 10^{54}$ erg (15 keV -- 5 MeV, see \citealp{2009PASJ...61..521S}).

Much of the progress made in observing GRBs in recent years has been due to the coordinated efforts of a large number of satellites including \emph{Konus}-WIND \citep{1995SSRv...71..265A}, \emph{Swift} \citep{2004ApJ...611.1005G}, \emph{Suzaku}-WAM \citep{2009PASJ...61S..35Y}, AGILE \citep{2009A&A...502..995T} and \emph{Fermi} \citep{2009ApJ...697.1071A,2009ApJ...702..791M}. These satellites allow an overall energy coverage from 0.2 keV to 300 GeV.

\begin{figure}
\centering
\includegraphics[width=\hsize,clip]{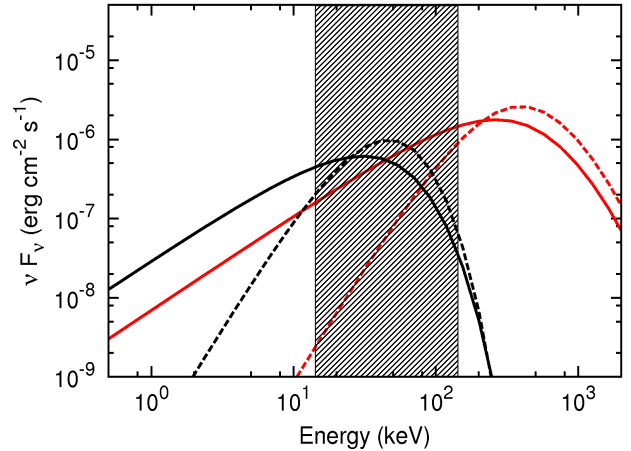}
\caption{Theoretically simulated spectra obtained by assuming two different values of $E_{iso}$: $10^{54}$ erg (in red) and $2.5\times 10^{53}$ erg (in black). Solid lines refer to the case $\alpha=-1.8$. Dotted lines refer to the case $\alpha=0.0$, corresponding to the pure thermal comoving spectrum (see Sec.~\ref{NewSed}). The box marks the energy band covered by \emph{Swift} BAT (15 keV -- 150 keV). The behaviour of the spectra around the peak energy is not significantly affected by the value of $\alpha$ (see Sec.~\ref{NewSed}). For the higher energetic sources, the low energy component of the spectra falls inside the energy band covered by BAT; this is not the case for the lower energetic sources, for which only the region around the peak can be investigated.}\label{fig:sp20s}
\end{figure}

GRB 080319B and GRB 050904 have been triggered by \emph{Swift}. What is relevant here is that, in both these sources with unusually high $E_{iso}$, the observed peak energy $E_{peak}^{obs}$ occurs well above the \emph{Swift} BAT bandpass (see Tab.~\ref{tab:epeak} in Sec.~\ref{sec:disc}): for GRB 050904 we have $E_{peak}^{obs}=314^{+173}_{-89}$ keV \citet{2009PASJ...61..521S}; for GRB 080319B we have $E_{peak}^{obs}=675 \pm 22$ keV. This, in turn, through the \emph{Swift} observations, for the first time allows the exploration of GRB spectra at $E \lesssim 0.1 E_{peak}$ (see Fig.~\ref{fig:sp20s}). This is at variance with the previous observations of lower energetics GRBs where $E_{peak}$ falls within the instrumental bandpass (see Fig.~\ref{fig:sp20s}). The observation of GRB 080319B has occurred prior to the launch of \emph{Fermi} and after the one of AGILE, but AGILE has been occulted by Earth during the burst detection. The high energy photons have been detected by \emph{Konus}-WIND \citep{2008Natur.455..183R}. The observation of GRB 050904 occurred prior to the launch of both AGILE and \emph{Fermi}. The high energy photons were detected by \emph{Suzaku}-WAM and \emph{Konus}-WIND \citep{2009PASJ...61..521S}.

GRB 080319B was discovered by BAT on March 19, 2008 \citep{2008Natur.455..183R}. It has a redshift $z=0.937$ \citep{2008GCN..7444....1V} and is characterized by an extraordinarily bright optical emission accompanying its $\gamma$-ray emission, that makes it the brightest optical burst ever observed: with a peak visual magnitude of 5.3, it could have been seen with the naked eye by an observer in a dark location \citep{2008Natur.455..183R}. It is also one of the brightest GRBs in $\gamma$ and X rays.

This extremely bright optical flash has promoted an intense theoretical analysis on the role of synchrotron self-Compton (SSC, see Sec.~\ref{sec:SEDprompt}) radiation and this scenario has been investigated by several authors, including e.g. \citet{2008Natur.455..183R} and \citet{2008MNRAS.391L..19K}.

There has been also interest in examining the possible existence of beaming in order to reduce the energetics of this source. \cite{2008Natur.455..183R} proposed a two-component jet model: a narrow jet, with a half-opening angle $\theta_{jet} \sim 0.2^\circ$, dominating the early emission, surrounded by a wider jet, with $\theta_{jet} \sim 4^\circ$, dominating the emission at intermediate and late times. This two-jet model would reduce the total energy budget to $\sim 4 \times 10^{50}$ erg. However, the narrow jet should produce a jet break $\sim$ 1 hr post-burst, which has not been observed. \citet{2008Natur.455..183R} suggested that this could be explained if the optical flux of the narrow jet is fainter than that of the wider jet. The wider jet should give rise to a break at $\sim 10^6$ s. \citet{2010ApJ...725..625T} reported the observation of a jet break at $\sim 11$ days with an ``approximately achromatic behaviour'', recalling that, in the \emph{Swift} era, the expected ``achromatic behaviour of breaks in X-ray and optical light curves has been rarely seen \citep{2008MNRAS.386..859C}''. An alternative single jet model has been considered by \citet{2008MNRAS.391L..19K}: taking into account the lack of an optical jet break during the first ten days of emission, they find a lower limit $\theta_{jet} \gtrsim 2^\circ$. The energy of the GRB could in this case be $\gtrsim 10^{52.3}$ erg (twice larger for a double-sided jet).

GRB 050904 was discovered by BAT on September 4, 2005 \citep{2005GCN..3910....1C}. It is one of the farthest GRBs ever observed, with $z=6.29$ \citep{2005GCN..3937....1K}. Also this burst is characterized by an intense optical emission: a bright optical flare was in fact detected by TAROT near the end of the prompt phase, in temporal coincidence with an X-ray flare \citep{2006ApJ...638L..71B,2007A&A...462..565G}.

\citet{2005A&A...443L...1T} found a ``steepening'' in the \emph{J}-band light curve at 2.6 $\pm$ 1.0 days and proposed that ``it may be due to a jet break''. In this case a jet opening angle $\theta_{jet}\sim 3^{\circ}$ could be inferred. This analysis was refined by \citet{2007AJ....133.1187K}, who put the steepening at 2.63 $\pm$ 0.37 days, corresponding to $\theta_{jet} \sim 3.30^\circ$. In this case the beaming-corrected energy is $1.73 \times 10^{51}$ erg \citep{2009PASJ...61..521S}.

In this paper we limit the analysis of these two sources to the $\gamma$-ray emission of the prompt phase, which is energetically predominant with respect to the optical emission: the optical isotropic energy is $\sim 0.1 \%$ of $E_{iso}$ \citep{2009arXiv0906.4144B} for GRB 080319B and is even less for GRB 050904 \citep{2006ApJ...638L..71B,2007A&A...462..565G}.

Also in view of the absence of achromatic breaks required by the jetted emission model, for GRB 080319B and GRB 050904 we assume spherical symmetry, which is one of the main features of the fireshell model.

We recall that a satisfactory agreement between the theoretical predictions of the fireshell model and the observed light curves and spectra of GRBs with $E_{iso}$ up to $10^{53}$ erg has been previously obtained \citep[see e.g.][]{2005ApJ...634L..29B,2006ApJ...645L.109R,2007A&A...471L..29D,2007A&A...474L..13B,2009A&A...498..501C,2010A&A...521A..80C,2011A&A...529A.130D}. What is really new in the analysis of these two sources is: 1) the very large value of $E_{peak}^{obs}$, well above the \emph{Swift} BAT bandpass (see Fig.~\ref{fig:sp20s} and Tab.~\ref{tab:epeak}); 2) the fact that for both sources the \emph{Swift} BAT bandpass covers the low energy part of the spectrum, that could then be investigated; 3) in the case of lower energetic sources, the energy region near the peak has been observed by BAT, while the low energy spectral component has been missed (see Fig.~\ref{fig:sp20s}) and will possibly be the object of future specific space missions.

In both GRB 080319B and GRB 050904, \emph{Swift} BAT data show more power in the low energy region than the theoretical predictions of the fireshell model (see Sec.~\ref{GRB080319B}). We then introduce a new phenomenological spectrum in the comoving frame of the fireshell: the ``modified'' thermal spectrum (see Sec.~\ref{NewSed}), characterized by a phenomenological parameter $\alpha$. The new comoving spectrum: 1) allows the correct reproduction of the observed prompt $\gamma$-ray emission light curves and spectra of both GRBs (see Secs.~\ref{GRB080319B} and \ref{GRB050904}); 2) clearly, as shown by Fig.~\ref{fig:sp20s}, does not modify significantly the considerations on the less energetic GRBs previously discussed within the fireshell model, since in that case only the region encompassing the peak is inside the instrument bandpass and is not significantly affected by the value of $\alpha$; 3) predicts a possible broader emission in the low energy spectral component of the lower energetics GRBs which will possibly be tested by future missions below 10 keV such as, e.g., LOFT \citep{2011arXiv1107.0436F} and MIRAX \citep{mirax}. These observations will determine the possible general validity of the comoving spectrum introduced here. It will also be important to verify if an analogous phenomenological correction in the Wien part of the comoving blackbody spectrum will allow the interpretation of the high energy data (above hundreds of MeV) later observed in equally energetic GRBs by AGILE and \emph{Fermi}. In this respect, it must be noted that an extra high-energy power-law component has been actually invoked in the current literature to explain the emission above hundreds MeV observed in e.g. GRB 090902B \citep{2009ApJ...706L.138A}, GRB 090510 \citep{2010ApJ...716.1178A} and GRB 090926A \citep{2011ApJ...729..114A}. However, the phenomenological extra power-law component characterizing the low energy tail of the new comoving spectrum introduced here is not necessarily related to the low energy extrapolation of such a high-energy component. In fact it is also present in sources which show no evidence of high energy emission, like e.g. GRB 090618 \citep{2012arXiv1202.4374I} and GRB 101023 \citep{2012A&A...538A..58P}.

The work is organized as follows. In Sec.~\ref{sec:SEDprompt} we describe the phenomenological and theoretical interpretation of GRB prompt emission. In Sec.~\ref{Fmodel} we briefly outline the fireshell model, we introduce the ``modified'' thermal spectrum and explain how we perform the numerical simulations to be compared with the observational data. In secs. \ref{GRB080319B} and \ref{GRB050904} we present, as specific examples, the analysis of GRB 080319B and GRB 050904 observations of the prompt $\gamma$-ray emission. In Sec.~\ref{sec:disc} we discuss the possible explanation for the need to introduce the ``modified'' thermal spectrum and in Sec.~\ref{concl} we present our conclusions.

\section{The spectral energy distribution of GRB prompt emission}\label{sec:SEDprompt}

The study of the GRB prompt emission is often performed by means of phenomenological approaches consisting, for example, in modeling the burst spectra with empirical functions whose parameters are fixed by the $\chi^2$ minimization procedure of the spectral data \citep[see, e.g., ][]{1998ApJ...496..849P,2000ApJS..126...19P,2002A&A...393..409G,2006ApJS..166..298K,2008ApJ...677.1168K,2011A&A...526A..49G}. The explanation of the results obtained from such phenomenological approaches are a main goal of the theoretical models of GRBs.

\subsection{Phenomenological approaches to GRB prompt emission}\label{phen}

The photon number spectrum time-integrated over the GRB prompt emission duration is typically best fit by two power-laws joined smoothly at a given break energy \citep[the ``Band function'', ][]{1993ApJ...413..281B}, whose low energy and high energy photon indices, $\alpha$ and $\beta$, have median values of $-1$ and $-2.3$ respectively \citep{2000ApJS..126...19P,2006ApJS..166..298K}.

Another example of phenomenological analysis is the correlation between the rest-frame value of the peak energy $E_{peak}^i$ and the isotropic equivalent radiated $\gamma$-ray energy $E_{iso}$, which leads to a power-law dependece: $E_{peak}^i \propto E_{iso}^{\lambda}$, with $\lambda \sim 0.57$ \citep{2006MNRAS.372..233A,2002A&A...390...81A,2009A&A...508..173A}.

There is also an additional phenomenological relation which has manifested itself in recent years, attracting attention: the possibility of fitting the observed spectra, integrated over selected time intervals, by a blackbody plus power-law (BB+PL). There are cases in which the GRB spectrum, again integrated over selected time intervals, is statistically indistinguishable between fits with a Band function and a BB+PL; there are other cases in which the BB+PL model is even preferred over the Band function and finally there are sources for which the data fulfill the Band function but not the BB+PL \citep{2004ApJ...614..827R,2005ApJ...625L..95R,2009ApJ...702.1211R,2010ApJ...709L.172R,2011MNRAS.tmp..935R}.

Generally, the additional blackbody component occurs in the early part of the prompt emission (see e.g. \citealp{2010ApJ...709L.172R}). There are bursts for which it is dominant with respect to the power-law, but opposite cases also exist, in which the power-law component is predominant with respect to the thermal one; the relative intensity of the two components can vary with time \citep{2004ApJ...614..827R,2005ApJ...625L..95R}.

In some interesting cases the blackbody component shows a characteristic evolution in time, with both its observed temperature and flux having a dependence from the arrival time well described by a broken power-law with indexes $a_T$, $b_T$, $a_F$ and $b_F$ \citep{2004ApJ...614..827R,2005ApJ...625L..95R,2009ApJ...702.1211R}. It is appropriate to emphasize that the correspondence between different spectral models cannot, in general, be applied to the entire GRB light curve. There is then the necessity of identifying the specific time interval over which the different spectral features are identified, also possibly in order to recover their physical origin \citep[see e.g.][]{COSPAR,TEXAS}.

Specifically, for GRB 080319B and GRB 050904 both time-resolved and time-integrated BAT spectra, including the \mbox{Proper-GRB} (P-GRB, see below Sec.~\ref{sec:canGRB}) spectrum of GRB 080319B, are best fit by a simple power-law (see Sec.~\ref{GRB080319B} and \ref{GRB050904}; see also \citealp{2008Natur.455..183R,2009AIPC.1133..356S} and \citealp{2007A&A...462...73C}); alternative models, such as a Band function or a BB+PL, cannot be constrained by the BAT data. No concluding statements on the presence of a blackbody component in the P-GRB of GRB 080319B can be made due to the limited bandpass of the instruments (see Sec.~\ref{GRB080319B}). The analysis of \emph{Konus}-WIND data from GRB 080319B has shown a best fit with a Band function \citep{2008Natur.455..183R}. For GRB 050904 \citet{2009PASJ...61..521S} performed a joint spectral analysis among \emph{Swift} BAT, \emph{Suzaku}-WAM and \emph{Konus}-WIND, finding a good fit with both a power-law with exponential cutoff and the Band function. A blackbody component has not been observed in either of these two sources. It is not clear at this stage if different conclusions could have been reached if these sources would have been observed by \emph{Fermi} or AGILE.

\subsection{Theoretical models of GRB prompt emission}

Many different models have been developed to theoretically explain the observational properties of GRBs. One of the most quoted ones is the fireball model (see \citealp{2004RvMP...76.1143P} for a review). An alternative one, originating in the electrodynamical processes around a Kerr-Newmann black hole \citep{1975PhRvL..35..463D}, is the fireshell model.

The fireball model was first proposed by \citet{1978MNRAS.183..359C}, \citet{1986ApJ...308L..47G} and \citet{1986ApJ...308L..43P}, who have shown that the release of a large quantity of $\gamma$-ray photons into a compact region can lead to an optically thick photon-lepton ``fireball'' through the production of $e^\pm$ pairs. The term ``fireball'' refers to an opaque plasma whose initial energy is significantly greater than its rest mass \citep{1999PhR...314..575P}.

The fireshell model also starts from an optically thick $e^\pm$ plasma whose evolution has been followed in a sequence of states of thermal equilibrium, taking properly into account the ultrarelativistic expansion and the detailed computation of the rate equation for the $e^\pm$ annihilation (see Sec.~\ref{sec:canGRB} and \citealp{1999A&A...350..334R,2000A&A...359..855R}).

In the fireball model, the prompt emission, including the sharp luminosity variations \citep{2000ApJ...539..712R}, are due to the prolonged and variable activity of the ``inner engine'' \citep{1994ApJ...430L..93R,2004RvMP...76.1143P}. The conversion of the fireball energy to radiation occurs via shocks, either internal (when faster moving matter overtakes a slower moving shell, see \citealp{1994ApJ...430L..93R}) or external (when the moving matter is slowed down by the external medium surrounding the burst, see \citealp{1992MNRAS.258P..41R}). Specifically, for GRB 080319B internal shocks have been considered responsible for the prompt emission (\citealp{2008Natur.455..183R}; see, however, \citealp{2009MNRAS.395..472K}) and external shocks are then considered responsible for the afterglow (see also, however, \citealp{1999ApJ...523L.113K} and references therein). For GRB 050904 \citet{2007A&A...462...73C} proposed that internal shocks are responsible for both the prompt and the afterglow emission (see, however, \citealp{2007A&A...462..565G} for a detailed description of limits and advantages of both the internal and the external shock scenario).

Much attention has been given to synchrotron emission from relativistic electrons, possibly accompanied by SSC emission to explain the observed GRB spectrum. These processes were found to be consistent with the observational data of many GRBs (see e.g. \citealp{1996ApJ...466..768T,2000ApJS..127...59F,1998ApJ...492..696S}). However, several limitations have been evidenced in relation with the low energy spectral slopes of time-integrated spectra (see \citealp{1997ApJ...479L..39C}, \citealp{2002ApJ...581.1248P}, \citealp{2002A&A...393..409G} and \citealp{2003A&A...406..879G}; see also, however, \citealp{2009arXiv0912.3743D}) and time-resolved spectra (see \citealp{1998AIPC..428..359C} and \citealp{2003A&A...406..879G}); additional limitations on SSC have also been pointed out by \citet{2008MNRAS.384...33K} and \citet{2009MNRAS.393.1107P}.

In all the above considerations, the equations of motion of the fireball are evaluated under the ultrarelativistic approximation, leading to the \citet{1976PhFl...19.1130B} self-similar power-law solution (see Sec.~\ref{sec:canGRB}). The maximum Lorentz factor of the fireball is estimated from the temporal occurrence of the peak of the optical emission, which is identified with the peak of the forward external shock emission \citep{2007A&A...469L..13M,2009ApJ...702..489R} in the thin shell approximation \citep{1999ApJ...520..641S}. It was also proposed to put an upper limit on the maximum fireball Lorentz gamma factor from the upper limit on the intensity of a possible smooth background signal in the hard X-rays to soft gamma rays during the prompt emission, which is identified with the contribution of the forward external shock emission to the prompt phase \citep{2010MNRAS.402.1854Z}. Another proposal was advanced to use compactness arguments within a scenario with two separate emitting regions for the MeV and the GeV emissions \citep{2011ApJ...726L...2Z}.

Partly alternative and/or complementary scenarios to the fireball model have been developed, e.g. the ones based on: quasi-thermal Comptonization \citep{1999A&AS..138..527G}, Compton drag emission \citep{1991ApJ...366..343Z,1994MNRAS.269.1112S,2000ApJ...529L..17L}, synchrotron emission from a decaying magnetic field \citep{2006ApJ...653..454P}, jitter radiation \citep{2000ApJ...540..704M}, Compton scattering of synchrotron self-absorbed photons \citep{2000ApJ...544L..17P,2004MNRAS.352L..35S}, photospheric emission \citep{2000ApJ...529..146E,2000ApJ...530..292M,2002ApJ...578..812M,2002MNRAS.336.1271D,2006A&A...457..763G,2009ApJ...702.1211R,2010ApJ...725.1137L}. In particular, \citet{2009ApJ...702.1211R} pointed out that photospheric emission overcomes some of the difficulties of pure non-thermal emission models.

In this paper we focus on the fireshell model. The characteristic parameters of the model are the total energy $E_{tot}^{e^\pm}$, the baryon loading $B$ and the CircumBurst Medium (CBM) density $n_{cbm}$ (see Secs.~\ref{sec:canGRB} and \ref{procedure}). The Lorentz $\gamma$ factor, directly linked to $B$ and $E_{tot}^{e^\pm}$, is explicitly computed from the description of all the phases starting from the moment of gravitational collapse. The fireshell radial coordinate is explicitly evaluated as a function of the laboratory time, the comoving time and the arrival time at the detector (see Sec.~\ref{sec:canGRB}). The optically thick $e^\pm$ plasma, endowed with baryon loading, is followed all the way to transparency. The collision with the CBM of the emerging relativistic expanding shell of baryons gives rise to the extended afterglow, which comprises both the prompt emission and the traditional decaying afterglow phases (see Sec.~\ref{sec:canGRB}). Relativistic effects are taken into account in the computation of the equations of motion of the shell and of the EQuiTemporal Surfaces \citep[EQTS, ][see Sec.~\ref{sec:easpectrum}]{2004ApJ...605L...1B,2005ApJ...620L..23B}. Taking into proper account these relativistic effects, it is possible to deduce the spectrum of the collision process between the baryons and the CBM in the comoving frame of the shell. For simplicity it was initially assumed that such collisions occur in a fully radiative regime and give rise to a pure thermal spectrum in the comoving frame (see Sec.~\ref{sec:easpectrum}). The observed spectrum is then obtained as a double convolution of thousands of thermal spectra, each one with a different temperature and weighted by the appropriate Lorentz and Doppler factors, following the solution of the equations of motion of the fireshell, both over the EQTS and over the observation time \citep{2004IJMPD..13..843R,2005ApJ...634L..29B}\footnote{Typically, a resolution of $\sim 5\times 10^4$ thermal spectra for each second of observation has been used.}.

As we recalled in the Introduction, this ansatz of a pure black body spectrum in the fireshell comoving frame, in spite of its simplicity, led to a successful interpretation of many different sources with an $E_{peak}^{obs}$ inside the instrumental bandpass \citep[see e.g.][]{2005ApJ...634L..29B,2006ApJ...645L.109R,2007A&A...471L..29D,2007A&A...474L..13B,2009A&A...498..501C,2010A&A...521A..80C,2011A&A...529A.130D}. We were then quite confident that we could obtain any observed power-law by an adequate convolution of thermal spectra duly weighted by the relativistic Doppler factors. The impossibility of obtaining the correct power-law indexes observed in both the sources treated in this paper, which have an $E_{peak}^{obs}$ above the instrumental bandpass, using solely thermal spectra in the fireshell comoving frame, was quite unexpected. This impossibility stems from the fact that neither the distribution of the temperature in the comoving frame, nor the Doppler factors used in the thousands of convolution processes, can be arbitrarily given. Both of them are constrained by the equations of motion of the fireshell and by the consequent release of kinetic energy in the collision. A priori, a convolution of thermal spectra with arbitrary temperatures and Doppler factors can always fit any observed power-law. There is, however, no way to fit by convolutions of thermal spectra the observed power-laws consistently with the fireshell equations of motion in these two sources with $E_{peak}^{obs}$ above the instrumental bandpass.

What is particularly interesting, however, is the fact that this difficulty can be overcome simply by adding an extra power-law component to the pure black body spectrum. As we will see, the application of the fireshell model to GRB 080319B and GRB 050904 leads in fact to the introduction of an additional phenomenological parameter $\alpha$, which characterizes the departure of the slope of the low energy part of the comoving spectrum from a pure thermal one (see Sec.~\ref{NewSed}).

This new result is consistent also with all the previously analyzed GRBs, and reasons for this are given in Fig.~\ref{fig:sp20s}. The success of this approach is not trivial, since there is no direct analytic relation between the index of the power-law introduced in the spectrum in the fireshell comoving frame and the observed one. It is quite significant that the introduction of a single power-law makes the fireshell model consistent with all observed GRB spectra.

The physical explanation for $\alpha$ has not yet been found. Analogously, no physical explanation has yet been found for the above described (Sec.~\ref{phen}) phenomenological parameters of the Band function and of the Amati relation, and in this sense for the ones described by \citet{2004ApJ...614..827R,2005ApJ...625L..95R} and \citet{2009ApJ...702.1211R} as well. All these phenomenological parameters lead to a better quantitative description of GRBs; they are the object of active theoretical studies and are an important step toward reaching the future identification of the underlying physical processes characterizing the GRBs.

We now proceed to a detailed description of the fireshell model (Sec.~\ref{Fmodel}).

\section{The fireshell model: the GRB luminosity and spectrum}\label{Fmodel}

The fireshell model, avoiding a piecewise fit of the observed GRB data, proposes a unified picture starting from the initial process of gravitational collapse to a black hole through all successive stages. It gives a theoretical treatment of each stage, identifying the parameters intrinsic to the source, essential to describing the evolution of the system, as well as its interaction with the CBM. The corresponding equations of motion are treated accordingly. Regimes of Lorentz gamma factors in the range $100$--$1000$ are encountered, implying the necessity of a fully relativistic approach. The model is intrinsically endowed with highly nonlinear effects: each prediction of the theoretical model at a given instant of the observation time is influenced by the entire history of the source. Consequently, any solution of the model in agreement with the observations necessarily implies a high level of self-consistency.

\subsection{The canonical GRB}\label{sec:canGRB}

Within the fireshell model, all GRBs originate from an optically thick electron--positron plasma in thermal equilibrium, having total energy $E_{tot}^{e^\pm}$ and formed in the gravitational collapse to a black hole \citep{2010PhR...487....1R}. The condition of thermal equilibrium assumed in our model and proved by \cite{2007PhRvL..99l5003A} distinguishes our model from alternative approaches \citep[e.g. the one by][]{1978MNRAS.183..359C}, where the total annihilation of the $e^\pm$ plasma was assumed, leading to a vast release of energy pushing on the CBM (the concept of a ``fireball''). In our case the annihilation of the $e^\pm$ pairs occurs gradually and is confined within an expanding shell: the ``fireshell''. The rate equation for the $e^\pm$ pairs and their dynamics (the pair-electromagnetic pulse or PEM pulse for short) has been given by \citet{1999A&A...350..334R}. This plasma engulfs the baryonic material left over in the process of gravitational collapse having mass $M_B$, still keeping thermal equilibrium between electrons, positrons and baryons. The baryon loading is measured by the dimensionless parameter $B=M_B c^2/E_{tot}^{e^\pm}$. It was shown \citep[see][]{2000A&A...359..855R} that no relativistic expansion of the plasma can be found for $B > 10^{-2}$. The fireshell is still optically thick and self-accelerates to ultrarelativistic velocities \citep[the pair-electromagnetic-baryonic pulse or PEMB pulse for short,][]{2000A&A...359..855R}. Then the fireshell becomes transparent and the P-GRB is emitted \citep{2001ApJ...555L.113R}. The amount of energy radiated in the P-GRB is only a fraction of $E_{tot}^{e^\pm}$. The remaining energy is stored in the kinetic energy of the optically thin baryonic and leptonic matter fireshell. The final Lorentz $\gamma$ factor at transparency, $\gamma_0$, can vary in a vast range between $10^2$ and $10^3$ as a function of $E_{tot}^{e^\pm}$ and $B$ \citep{2000A&A...359..855R}.

After transparency, the remaining accelerated baryonic matter still expands ballistically and starts to slow down by the collisions with the CBM, having average density $n_{cbm}$. During this phase, the extended afterglow emission occurs \citep{2001ApJ...555L.113R}. In common with the majority of existing models, we describe the motion of the baryons as an expanding thin shell enforcing energy and momentum conservation in the collision with the CBM. The condition of a fully radiative regime is assumed \citep{2003AIPC..668...16R}. It is appropriate to recall a further difference between our treatment and the ones in the current literature. The complete analytic solution of the equations of motion of the baryonic shell has been developed \citep{2004ApJ...605L...1B,2005ApJ...620L..23B}, while in the current literature usually the \citet{1976PhFl...19.1130B} self-similar solution has been uncritically adopted \citep[e.g.][]{1993ApJ...415..181M,1997ApJ...489L..37S,1998ApJ...494L..49S,1997ApJ...491L..19W,1998ApJ...496L...1R,1999ApJ...513..679G,1998ApJ...493L..31P,1999PhR...314..575P,1999ApJ...511..852G,2000ARA&A..38..379V,2002ARA&A..40..137M}. The similarities and differences between the two approaches have been explicitly pointed out in \citet{2005ApJ...633L..13B}.

From this general approach, a canonical GRB bolometric light curve is defined which is composed of two different parts: the P-GRB and the extended afterglow. The relative energetics of these two components, as well as the observed temporal separation between the corresponding peaks, is a function of the above three parameters $E_{tot}^{e^\pm}$, $B$, and $n_{cbm}$; the first two parameters are inherent to the accelerator characterizing the GRB, i.e., the optically thick phase, while the third one is inherent to the GRB surrounding environment which gives rise to the extended afterglow. What is usually called the GRB ``prompt emission'' in the literature is actually composed of both the P-GRB and the initial part of the extended afterglow encompassing its peak. As we proposed in \citet{2001ApJ...555L.113R}, both the so-called ``short'' and ``long'' GRBs fit into this canonical GRB scenario. In particular, for baryon loading $B \lesssim 10^{-5}$, the P-GRB component is energetically dominant over the extended afterglow. In the limit $B \to 0$ it gives rise to a ``genuine short'' GRB. Otherwise, when $3.0\times 10^{-4} \lesssim B \leq 10^{-2}$, the kinetic energy of the baryonic and leptonic matter, and consequently the extended afterglow emission, predominates with respect to the P-GRB, giving rise to the ``long'' GRBs or the ``disguised short'' GRBs depending on the average CBM density and the astrophysical scenario \citep{2001ApJ...555L.113R,2002ApJ...581L..19R,2007A&A...474L..13B,2009A&A...498..501C,2010A&A...521A..80C,2011A&A...529A.130D}. Since the ``critical'' value of $B$ (i.e. the value of $B$ for which both the P-GRB and the extended afterglow have the same energy) is a slowly varying function of $E_{tot}^{e^\pm}$, for $10^{-5}\lesssim B \lesssim 3.0\times 10^{-4}$ the ratio of the total energies of the P-GRB and of the extended afterglow is also a function of $E_{tot}^{e^\pm}$ \citep{2009AIPC.1132..199R}.

If one goes to the observational properties of this model of a relativistic expanding shell, a crucial concept has been the introduction of the EQTS. In this topic, also, our model is distinguished from those in the literature for deriving analytic expressions for the EQTS from the analytic solutions of the equations of motion \citep{2005ApJ...620L..23B}.

The observed temporal variability of the extended afterglow is produced in our model by the interaction of the fireshell with CBM ``clumps'' \citep[][see also Sec.~\ref{procedure}]{2002ApJ...581L..19R}. The issue of time variability in GRB light curves has been longly debated. Several authors, e.g. \citet{2006ApJ...642..354Z,2007MNRAS.380.1744N}, found that CBM inhomogeneities are not able to produce the short-timescale variability in GRB prompt and afterglow emission light curves, mainly because photons emitted at the same instant of time from different parts of the emitting regions have different arrival times: this tends to smoothen the light curve significantly.

Within the fireshell model, also on this point there are some differences with the other models. It is emphasized that, from the correct computations of the equations of motion of the shell and of the Lorentz $\gamma$ factor, the short time scale variability of GRB light curves occurs in regimes with the largest values of the Lorentz gamma factor, when the total visible area of the emission region is very small and the ``dispersion'' in arrival time of the luminosity peaks is negligible. Therefore, under this condition the short-timescale variability of GRB light curves can be produced by inhomogeneities in the CBM, as found also by \citet{1999ApJ...513L...5D} and \citet{2006NCimB.121.1331D,2008ApJ...684..430D}. The application of the fireshell model leads to a direct evaluation of the filamentary and clumpy structure of the CBM, which was already predicted in pioneering works by Enrico Fermi in the theoretical study of Interstellar Matter (ISM) in our galaxy \citep{1949PhRv...75.1169F,1954ApJ...119....1F}, and is much on line with the knowledge obtained from various studies of the ISM in galaxies \citep[see, for example,][]{1998PASA...15..132K,2002ApJ...580L..47L,2009ApJ...691..465F}.

Clearly, the fact that the short time scale variability is observed only in the prompt emission, while the X-ray afterglow light curves are usually smooth, does not mean that the CBM is inhomogeneous only up to a given radius, beyond which it becomes homogeneous. Inhomogeneities are everywhere in the CBM, but beyond a given distance from the source, corresponding approximately to the end of the prompt emission phase, they do not produce observable effects on the light curve since they are indeed smeared out by the curvature effect and the relativistic effects between the time in the fireshell comoving frame and the photon arrival time at the detector \citep{2002ApJ...581L..19R,2006ApJ...645L.109R}. The early X-ray afterglow originates in fact from the same kind of interaction of the fireshell with a clumpy CBM. The absence of spiky emission is simply a consequence of these effects. In other words, it is true that the fireshell model predicts that the same clumps at larger radii would produce longer spikes \citep{2001NCimB.116...99R}. However, at such large radii where the effect would be measurable, the smearing dominates and prevents the effect to be observed \citep{2002ApJ...581L..19R,2006ApJ...645L.109R}. Vice versa, the prompt emission, where the effect of the CBM clumps are observable, occurs encompassing too limited a range of radial distances to make this effect noticeable in all sources. The only exception may occur in the case of isolated high density clumps along the line of sight at late time of emission where indeed this effect is observable \citep{2006NCimB.121.1441B,2008AIPC.1065..227B, 2009AIPC.1111..383B,2009AIPC.1132..199R,2010JKPS...57..551I}. We are currently verifying if this aspect of the fireshell model may also explain some specific properties of the X-ray flares recently evidenced in the afterglow phase \citep[see e.g.][and references therein]{2011MNRAS.410.1064M,2011MNRAS.417.2144M}. The drop in energy of 2--4 orders of magnitude from the prompt $\gamma$-ray phase to the early X-ray afterglow is perfectly in line with the decrease of the Lorentz gamma factor in the expansion of the fireshell. The fact that the prompt emission in $\gamma$-rays stops at some time is related to the well known hard-to-soft evolution of the emission process, which is perfectly explained within the fireshell model (see Sec.~\ref{sec:easpectrum}) and it is related to the solution of the equations of motion of the system, implying the decrease of the Lorentz gamma factor as well as the amount of energy release in the collision between the fireshell baryonic matter and the CBM \citep{2004IJMPD..13..843R,2005ApJ...634L..29B}.

\subsection{The P-GRB spectral properties}\label{sec:pgrb}

The spectrum at transparency has been given in \citet{2000A&A...359..855R} with a temperature computed consistently with the local thermodynamics of the $e^\pm$-baryon plasma and the Lorentz gamma factor. Details are given in \citet{2009AIPC.1132..199R}. It is appropriate to stress that in the emission of the P-GRB there are two different contributions: one corresponding to the emission of the photons due to the reach of the transparency, and the second originating from the interaction of the protons and electrons with the CBM. A spectral energy distribution with a thermal component and a non-thermal one should be expected to occur.

\subsection{The extended afterglow spectral properties}\label{sec:easpectrum}

The majority of work in the current literature has addressed the analysis of the prompt emission as originating from various combinations of synchrotron and inverse Compton processes \citep{2004RvMP...76.1143P}. It appears clear, however, that this interpretation is not satisfactory (see Sec.~\ref{intro}; see also \citealp{2003A&A...406..879G,2008MNRAS.384...33K,2009MNRAS.393.1107P}). Furthermore, in the description of an ultrarelativistic collision between protons and electrons and the CBM new collective processes of ultrarelativistic plasma physics occur, not yet fully explored and understood \citep[e.g. Weibel instability, see][]{1999ApJ...526..697M}. Most promising results along this line have been already obtained by \citet{2008ApJ...673L..39S,2009ApJ...700..956M}.

Without waiting for the developments of these investigations, we have adopted a very pragmatic approach in the fireshell model by making full use of the knowledge of the equations of motion, of all the EQTS formulations as well as of the correct relativistic transformations between the comoving frame of the fireshell and the observer frame. In this respect, we have adopted a fundamental procedure: to make an ansatz on the spectral properties of emission of the collisions between the baryons and the CBM in the comoving frame, and then evaluate all the observational properties in the observer frame. In order to take into proper account the filamentary, clumpy and porosity structure of the CBM, we have introduced an additional parameter ${\cal R}$, which describes the fireshell surface filling factor. It is defined as the ratio between the effective emitting area of the fireshell $A_{eff}$ and its total visible area $A_{vis}$ \citep{2002ApJ...581L..19R,2004IJMPD..13..843R,2005IJMPD..14...97R}.

It must be emphasized that the fact that only a fraction ${\cal R}$ of the shell surface is emitting does not mean that only a fraction ${\cal R}$ of the total shell energy is emitted. We must in fact distinguish between an instantaneous interaction of the fireshell with a single filament and its overall interaction with all the filaments of the entire cloud giving rise to the spiky structure of the light curve. This global interaction is clearly the superposition of randomly distributed instantaneous events. The different filaments inside the cloud interact with different parts of the fireshell and the entire cloud reduces the kinetic energy of the entire fireshell. The key point is that, during the prompt emission, the cloud, typically with a mass of the order of $10^{-8}-10^{-11}$ solar masses, covers the entire visible area of the fireshell (typically with a radius between $10^{13}$ cm and $10^{15}$ cm for the sources analyzed in the
present paper, see Tables \ref{tab:08density} and \ref{tab:05density}). Consequently, at any given instant of time, each filament of the cloud covers only a small fraction of the fireshell surface. However, when we integrate over the cloud crossing time, the coverage of the cloud as a whole is equal to unity.

As a first ansatz, we have assumed that the extended afterglow radiation has a thermal spectrum in the comoving frame of the fireshell \citep{2004IJMPD..13..843R}. The observed GRB spectrum is given by the convolution of hundreds of thermal spectra with different temperatures and different Lorentz and Doppler factors. Such a convolution is to be performed over the EQTSs \citep{2004ApJ...605L...1B,2005ApJ...620L..23B}, which are the surfaces of constant arrival time of the photons at the detector, and over the observation time \citep{2005ApJ...634L..29B}.

The hard-to-soft transition of GRB time-integrated and time-resolved spectra, that was first observed in BATSE GRBs \citep{1997ApJ...479L..39C}, within the fireshell model comes out naturally from: 1) the evolution of the comoving temperature; 2) the decrease of the bulk $\gamma$ factor; 3) the curvature effect \citep{2004ApJ...605L...1B,2004IJMPD..13..843R,2005ApJ...634L..29B}.

Within the fireshell model, the extended afterglow luminosity at the detector arrival time $t_a^d$ per unit of solid angle $d\Omega$ and in the energy band $[\nu_1,\nu_2]$ is given by \citep{2004IJMPD..13..843R}
\begin{equation}\label{eq:SL}
\frac{dE^{[\nu_1,\nu_2]}}{dt_a^d d\Omega}=\int_{EQTS} \frac{\Delta \epsilon}{4 \pi} v \cos\theta \Lambda^{4} \frac{dt}{dt_a^d} W(\nu_1,\nu_2,T_{arr}) d\Sigma,
\end{equation}
where $\Delta\epsilon=\Delta E_{int}/V$ is the emitted energy density released in the interaction of the accelerated baryons with the CBM measured in the comoving frame, $\Lambda=\left\{\gamma [1-(v/c) cos\theta]\right\}^{-1}$ is the Doppler factor, W($\nu_1$,$\nu_2$,$T_{arr}$) is an ``effective weight'' required to evaluate only the contributions in the energy band $[\nu_1,\nu_2]$, $d\Sigma$ is the surface element of the EQTS at detector arrival time $t_a^d$ on which the integration is performed and $T_{arr}$ is the observed temperature of the radiation emitted from $d\Sigma$. The ``effective weight'' W($\nu_1$,$\nu_2$,$T_{arr}$) is defined as the ratio between the energy density emitted in a given energy band $[\nu_1,\nu_2]$ and the bolometric energy density:
\begin{equation}\label{eq:wf}
W(\nu_1,\nu_2,T_{arr})=\frac{\int_{\epsilon_1}^{\epsilon_2} \left(\frac{dN_\gamma}{dV d\epsilon}\right) \epsilon d\epsilon }{\int_0^\infty \left(\frac{dN_\gamma}{dV d\epsilon}\right) \epsilon d\epsilon},
\end{equation}
where $\frac{dN_\gamma}{dV d\epsilon}$ is the number density of the photons per unit of energy in the comoving frame of the fireshell. 

\subsubsection{Thermal case}\label{thermal}

With the assumption of a thermal spectrum in the comoving frame of the fireshell 
\begin{equation}\label{eq:denfth}
\frac{dN_\gamma}{dV d\epsilon}=\left(\frac{8 \pi}{h^3 c^3}\right) \frac{\epsilon^2}{\exp\left(\frac{\epsilon}{k_B T}\right)-1},
\end{equation}
($h$ is the Planck constant, $c$ is the speed of light and $k_B$ is the Boltzmann constant) we have
\begin{equation}
W(\nu_1,\nu_2,T_{arr})=\frac{\int_{\epsilon_1}^{\epsilon_2} \left(\frac{dN_\gamma}{dV d\epsilon}\right) \epsilon d\epsilon}{a T^4},
\end{equation} 
where $a$ is the radiation constant and $T$ is the temperature in the comoving frame. 

In general, the temperature in the comoving frame can be evaluated starting from the following relation:
\begin{equation}\label{eq:calT}
\frac{\Delta E_{int}}{\Delta \tau}= \pi r^2 c\,{\cal R} \int_{0}^{\infty} \left(\frac{dN_\gamma}{dV d\epsilon} \right) \epsilon d\epsilon,
\end{equation}
where $\Delta \tau$ is the time interval in which the energy $\Delta E_{int}$ is developed. By inserting Eq.~(\ref{eq:denfth}) into Eq.~(\ref{eq:calT}) we obtain
\begin{equation}\label{eq:Tth}
T=\left(\frac{\Delta E_{int}}{4 \pi r^2 \sigma {\cal R} \Delta \tau}\right)^{1/4},
\end{equation}
with $\sigma$ the Stefan-Boltzmann constant.

\subsubsection{The ``modified'' thermal spectrum}\label{NewSed}

The new SED for the radiation emitted in the comoving frame of the fireshell is a ``modified'' thermal spectrum: a spectrum characterized by a different asymptotic power-law index in the low energy region with respect to the thermal one. This index is represented by a free parameter $\alpha$, so that the pure thermal spectrum corresponds to the case $\alpha=0$:
\begin{equation}\label{eq:denfmod}
\frac{dN_\gamma}{dV d\epsilon}=\left(\frac{8 \pi}{h^3 c^3}\right) \left( \frac{\epsilon}{k_B T}\right)^{\alpha} \frac{\epsilon^2}{\exp\left(\frac{\epsilon}{k_B T}\right)-1}.
\end{equation}
$\alpha$ is a phenomenological parameter defined in the comoving frame of the fireshell. This phenomenological approach can be relevant in identifying the true physical mechanisms occurring in the collisions in the comoving frame, and uniquely separates them from the relativistic contributions coming from relativistic transformations and convolutions over the EQTS leading to the observed spectrum.

By using the Eq.~(\ref{eq:denfmod}) and introducing the variable $y=\epsilon/(k_B T)$, we obtain the following expression for the ``effective weight'':
\begin{equation}
W(\nu_1,\nu_2,T_{arr})=\frac{\int_{y_1}^{y_2} \frac{y^{\alpha+3}}{\exp(y)-1} dy}{\Gamma(4+\alpha)\,Li_{4+\alpha}(1)}
\end{equation}
where $\Gamma(z)=\int_0^\infty t^{z-1} e^{-t} dt$ is the Gamma function and \mbox{$Li_n(z)=\sum_{k=1}^{\infty} z^k/k^n$} is Jonqui\`ere's function.

In analogy with the thermal case, we can define an ``effective temperature'' for the ``modified'' thermal spectrum; by inserting Eq.~(\ref{eq:denfmod}) in Eq.~(\ref{eq:calT}) we obtain:
\begin{equation}\label{eq:Tmodth}
T=\left[
\left(\frac{\Delta E_{int}}{\Delta \tau}\right) \frac{h^3 c^2}{(4 \pi r^2) \, 2 \pi {\cal R} \, k_B^4 \, \Gamma(4+\alpha)\,Li_{4+\alpha}(1)}
 \right]^{1/4}.
\end{equation}
It can be easily seen that, for $\alpha=0$, we obtain Eq.~(\ref{eq:Tth}). 

\begin{figure}
\centering
\includegraphics[width=\hsize,clip]{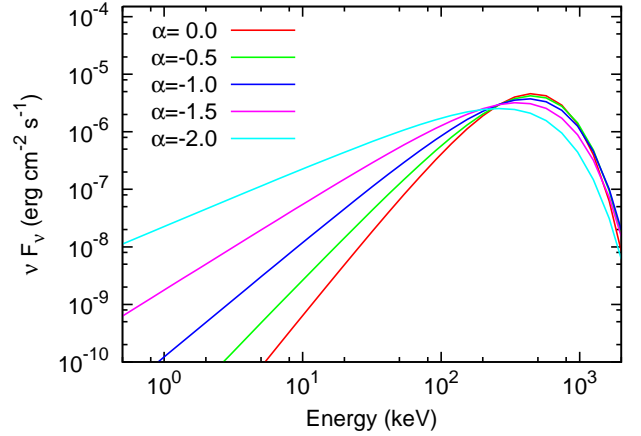}
\caption{Theoretically simulated instantaneous spectra obtained by assuming $E^{tot}_{e^+ e^-}=1.0\times10^{54}$ erg, $B=2.5\times 10^{-3}$, $n_{cbm}=1$ part cm$^{-3}$ and different values of the index $\alpha$. The curve with $\alpha=0.0$ corresponds to the pure thermal spectrum case.}\label{fig:spectra}
\end{figure}

In Fig.~\ref{fig:spectra} are shown several theoretically predicted instantaneous spectra characterized by the same temperature and different values of the free parameter $\alpha$. It can be seen that the main effect of varying the value of $\alpha$ is a change in the low energy slope of the spectral energy distribution. In particular, by decreasing $\alpha$ the low energy emission increases. Around the peak energy the spectrum is instead only weakly dependent on the value of $\alpha$. 

In the present paper we test the comoving modified thermal spectrum by comparing the numerical simulations with the observed prompt emission spectra and light curves of highly energetic GRBs. In particular we present, as specific examples, the analysis of the observational BAT data (in the $15$--$150$ keV bandpass) of GRB 080319B ($E_{iso}=1.34 \times 10^{54}$ erg, \citealp{2008GCN..7482....1G}) and GRB 050904 ($E_{iso}=1.04^{+0.25}_{-0.17}\times 10^{54}$ erg, \citealp{2009PASJ...61..521S}) in Secs.~\ref{GRB080319B} and \ref{GRB050904}, respectively.

\subsection{The numerical simulation of GRB light curves and spectra}\label{procedure}

To best reproduce the observational data within the fireshell model, we need to determine the following five parameters: $E_{tot}^{e^\pm}$, $B$, $\alpha$, $n_{cbm}$ and ${\cal R}$.

The procedure assumes a specific value of $E_{tot}^{e^\pm}$ and $B$. It is clear that $E_{tot}^{e^\pm}$ has to be larger or equal to the observed isotropic equivalent energy $E_{iso}$ of the GRB. $E_{tot}^{e^\pm}$ can be actually quite larger than $E_{iso}$ since, in many sources, we are limited by the threshold and the bandpass of the detectors. The value of $B$ is determined by the ratio between the energetics of the P-GRB and of the extended afterglow, as well as by the time separation between the corresponding peaks \citep{2001ApJ...555L.113R,2008AIPC.1065..219R,aksenovVenice}.

The determination of the three remaining parameters depends on the detailed ``fitting'' of the shape of the extended afterglow light curves and spectra. In particular, the parameter ${\cal R}$ determines the effective temperature in the comoving frame and the corresponding peak energy of the spectrum, $\alpha$ determines the low energy slope of the comoving spectrum and $n_{cbm}$ determines the temporal behavior of the light curve. It is found that the CBM is typically formed of ``clumps'' of width $\sim 10^{15} - 10^{16}$ cm and density contrast $10^{-1} \lesssim \delta n/n \lesssim 10$. Particularly important is the determination of the average value of $n_{cbm}$. Values of the order of $0.1$--$10$ particles/cm$^3$ have been found for GRBs exploding inside star forming region galaxies, while values of the order of $10^{-3}$ particles/cm$^3$ have been found for GRBs exploding in galactic halos \citep[i.e. the ``disguised'' GRBs, see ][]{2007A&A...474L..13B,2009A&A...498..501C,2010A&A...521A..80C,2011A&A...529A.130D}.

Of course, ``fitting'' a GRB within the fireshell model is much more complex than simply fitting the $N(E)$ spectrum with phenomenological analytic formulas for a finite temporal range of the data. It is a consistent picture, which has to ``fit'' the intrinsic parameters of the source, as well as its spectrum and its light curve temporal structure. Concerning the theoretical spectrum to be compared with the observational data, it is obtained by an averaging procedure of instantaneous spectra. In turn, each instantaneous spectrum is linked to the fit of the observed multiband light curves in the chosen time interval. Therefore, both the ``fit'' of the spectrum and of the observed multiband light curves have to be performed together and jointly optimized. Moreover, the parameters used in the numerical simulations are not independent. In fact, they have to be computed self-consistently through the entire dynamical evolution of the system and not separately at each time step. For each spike in the light curve the parameters of the corresponding CBM clumps must be computed, taking into proper account all the thousands of convolutions of comoving spectra over each EQTS leading to the observed spectrum. It is clear then that since the EQTS encompass emission processes occurring at different comoving times, weighted by their Lorentz and Doppler factors, the ``fitting'' of a single spike of the light curve is not only a function of the properties of the specific CBM clump but of the entire previous history of the source. Any step of the ``fitting'' process affects the entire following evolution and, viceversa, at any step a ``fit'' must be made consistently with all the previous and subsequent history: due to the non linearity of the system and to the EQTS, any change in the ``fit'' produces observable effects up to a much later time. This implies that the ``fitting'' process cannot proceed for successive temporal steps: the complete analysis must be applied to the entire GRB as a whole, to avoid possible systematic error propagation from a temporal step to the following ones. This leads to an extremely complex trial and error procedure in the fitting of the data in which the uniqueness of the parameters defining the source are further and further narrowed down. Of course, we cannot expect the latest parts of the fit to be very accurate, since some of the basic hypotheses on the equations of motion, and the possible fragmentation of the shell \citep{2007A&A...471L..29D}, can affect the fitting procedure.

\section{GRB 080319B}\label{GRB080319B}

We analyzed the GRB 080319B prompt emission light curve and spectrum observed by BAT within the fireshell model. As we already mentioned in the introduction, for GRB 080319B we have $E_{peak}^{obs}=675 \pm 22$ keV, although there is a hard-to-soft spectral evolution going from with $E_{peak}^{obs}=(748 \pm 26)$ keV at $22$ s after the BAT trigger to $E_{peak}^{obs}=(528 \pm 28)$ keV at $24$ s after the BAT trigger, and the BAT spectral index reduces from $\sim 1.0$ to $\sim 2.1$ at $53$ seconds after the BAT trigger \citep{2008Natur.455..183R}.

Several authors found some evidence of the possibility to separate the prompt emission of this source into two main episodes, partitioned at about 28 s after the BAT trigger time. \citet{2008AIPC.1065..259M} analyzed the variability time-scale $t_{var}$ of the $\gamma$-ray prompt emission, finding that the first part of the light curve (up to $\sim$ 28 s) is dominated by $t_{var} \sim$ 0.1 s, while the last part shows a much longer characteristic time-scale ($t_{var} \sim$ 0.7 s). \citet{2009AIPC.1133..356S} found that the arrival offset between the Swift-BAT 15-25 keV and 50-100 keV energy band ($\gamma$-ray spectral lag) is maximum at t $\gtrsim$ 28 s and it appears to be anti-correlated with the arrival offset between prompt 15-350 keV $\gamma$-rays and the optical emission observed by TORTORA (optical/$\gamma$-ray spectral lag), maximum at t $\lesssim$ 28 s.

\begin{figure}
\centering
\includegraphics[width=\hsize,clip]{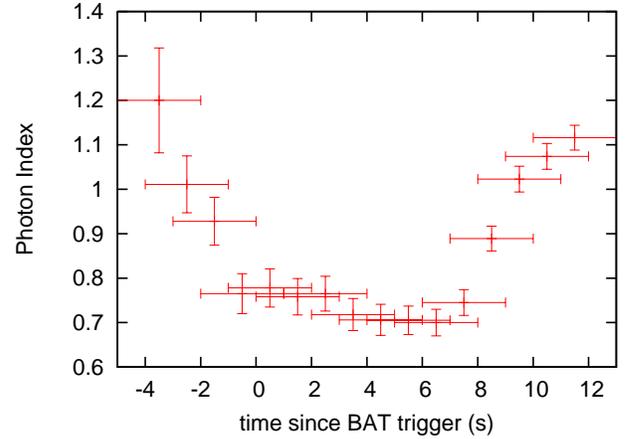}
\caption{Time evolution of the photon index for power-law fit to 15-150 keV Swift BAT spectra integrated over time intervals of 3 s. A clear soft-to-hard evolution up to $\sim$ 1 s after the BAT trigger time, while a typical hard-to-soft transition starts at about 8 s after the BAT trigger time. In the region between $\sim$ 1 s and $\sim$ 8 s, the photon index first appears to reach an asymptotic value of $\sim$ 0.76, then further decreases up to $\sim$ 0.7.}\label{fig:PhIndex}
\end{figure}

\begin{figure}
\centering
\includegraphics[width=\hsize,clip]{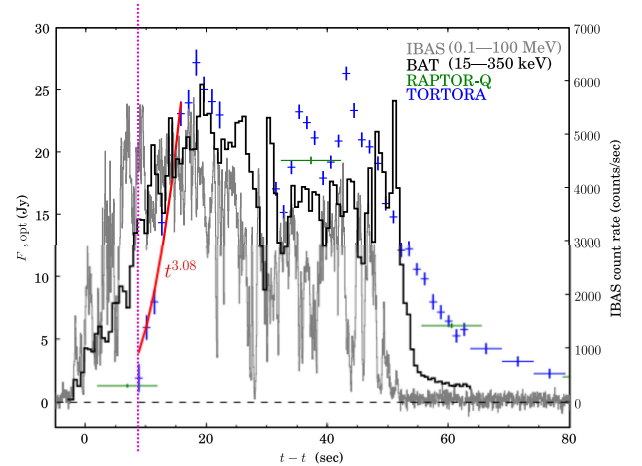}
\caption{Comparison of the prompt optical light curve (TORTORA, blue points; RAPTOR-Q, green points; Pi of the Sky, blue points) and the hard X-ray to $\gamma$-ray light curve (IBAS, grey points; BAT, black points) of GRB 080319B; the pink dotted line marks the begin of the optical emission detected by TORTORA (t $\sim$ 9 s after the BAT trigger time), whose onset is very rapid ($\sim$ t$^{3.08}$). Reproduced from \citet{2009ApJ...691..495W} with kind permission of P.R. Wo{\'z}niak and of AAS.}\label{fig:Optical}
\end{figure}

Concerning the first episode, we identify the first 7 s of emission (from -5 s up to 2 s after the BAT trigger time) with the P-GRB; the theoretically estimated total isotropic energy emitted in the P-GRB and the observed temperature are $E^{iso}_{P-GRB}=1.85\times10^{52}$ erg and $T^{obs}_{P-GRB}$ $\sim$ 16 keV respectively. There are three main reasons that supports this interpretation:
\begin{enumerate}
\item First, we performed the analysis of BAT spectra integrated over sub-intervals of time encompassing the whole prompt emission by using the standard FTOOLS package (Heasoft, version 6.10). We found that all these spectra are best modeled with power-laws and a discontinuity in the hard-to-soft evolution came out a few seconds after the BAT trigger time, as shown in Fig.~\ref{fig:PhIndex} \citep[see also][]{2009AIPC.1133..356S}. In particular, there is a clear soft-to-hard evolution up to $\sim$ 1 s after the BAT trigger time, while a typical hard-to-soft transition starts at about 8 s after BAT trigger time. It is difficult to evaluate at what time exactly the discontinuity occurs: in fact, in the region between $\sim$ 1 s and $\sim$ 8 s, the photon index first appears to reach an asymptotic value of $\sim$ 0.76, then further decreases up to $\sim$ 0.7: this behaviour could be due to the partial superimposition of the contributions of both the P-GRB and the extended afterglow.
\item The second reason suggesting the interpretation of the first seconds of emission as the P-GRB is that the optical emission starts at t $\sim$ 9 s after the BAT trigger time (see Fig.~\ref{fig:Optical}): in fact, within the fireshell model the optical radiation is expected in the extended afterglow, but not in the P-GRB.
\item It is important to point out that, besides the two above observational considerations, the identification of the first 7 s of emission as the P-GRB is the only interpretation that allow us to constrain the values of $E_{tot}^{e^\pm}$ and $B$ in such a way as to make a consistent fit of both the P-GRB and the extended afterglow.
\end{enumerate}
Of course, none of these three arguments separately would give support to our interpretation: e.g., the optical component could have also been present before $\sim 9$ s but unobserved due to the instrumental constraints. However, the redundant occurrence of all three arguments implies that our interpretation of the first 7 s as the P-GRB component is fully compatible both with our theoretical framework and with the observed data. The P-GRB spectrum is expected to be composed of a thermal plus a non-thermal component (see Sec.~\ref{sec:pgrb}). We found that the BAT spectrum of the first 7 s of the prompt emission is best fit by a power-law with photon index $\gamma$=0.84 $\pm$ 0.04, with a chi square value of $\chi^2=48.26$ for 60 degrees of freedom. The presence of the expected additional thermal component cannot be constrained by the data (C. Guidorzi, private communication). A possibility is that thermal flux is much lower than the non-thermal one and then it is negligible. Another alternative and/or complementary possibility is that the thermal component has been missed due to the limited bandpass of the instruments. It can be only matter of speculation if observations by \emph{Fermi} and AGILE would have led to a different conclusion.

\begin{figure}
\centering
\includegraphics[width=\hsize,clip]{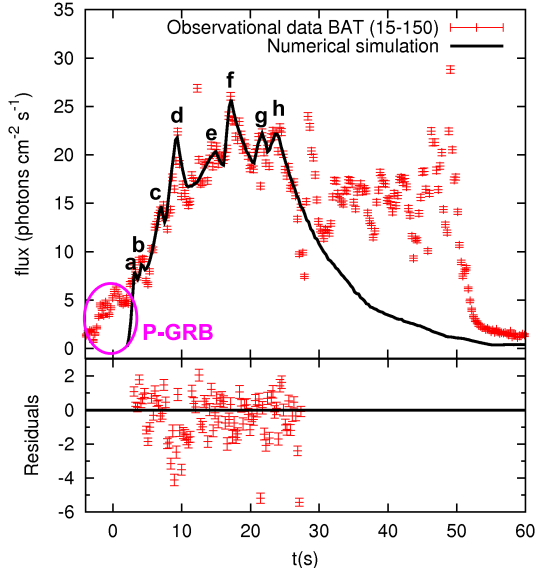}
\caption{Theoretically simulated light curve of GRB 080319B prompt emission in the 15-150 keV energy band (black solid curve) is compared with the data observed by BAT (red points); the P-GRB is marked with a magenta circle. The vertical dotted line marks the begin of the second part of the prompt emission (t $\sim$ 28 s). The labels ``a'', ``b'', ``c'', ``d'', ``e'', ``f'', ``g'' and ``h'' identify the peaks (see Fig.~\ref{fig:08density} and Tab.~\ref{tab:08density}).}\label{fig:LC08}
\end{figure}

\begin{figure}
\centering
\includegraphics[width=\hsize,clip]{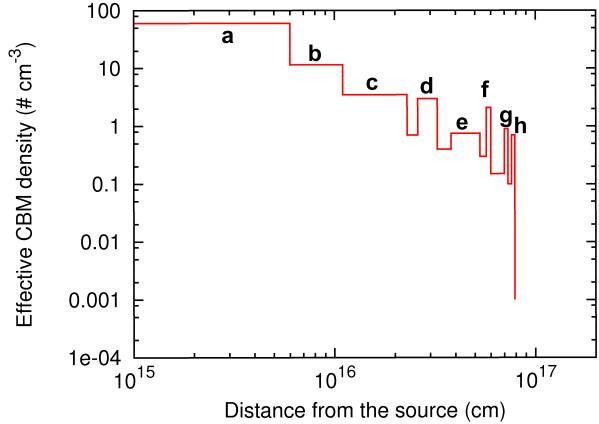}
\caption{Structure of the CBM adopted for GRB 080319B. The labels ``a'', ``b'', ``c'', ``d'', ``e'', ``f'', ``g'' and ``h'' indicate the values corresponding to the peaks in the BAT light curve (see Fig.~\ref{fig:LC08} and Tab.~\ref{tab:08density}).}\label{fig:08density}
\end{figure} 

\begin{figure}
\centering
\includegraphics[width=\hsize,clip]{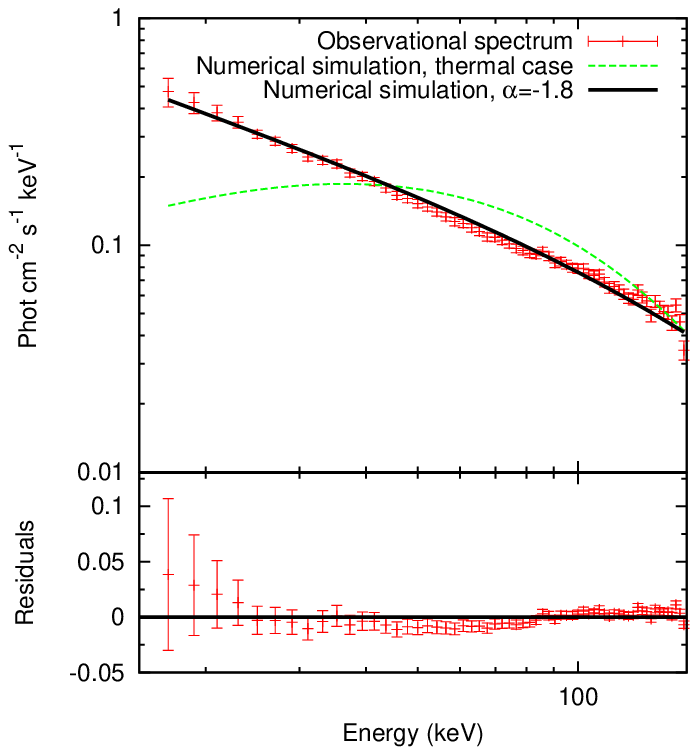}
\caption{Theoretically simulated spectra of GRB 080319B integrated over the time interval $3 s \leq t_a^d \leq 28 s$ with $\alpha=-1.8$ (black solid line) and $\alpha=0.0$ (pure thermal case, green dashed line) are compared with the data observed by BAT (red points). It can be seen that with the ``modified'' thermal spectrum we can correctly reproduce the observed spectrum, contrary to what happens with the pure thermal spectrum. In the residual plot the pure thermal case is omitted.}\label{fig:spectrum08}
\end{figure}

The remaining part of the first episode (from $\sim$ 7 s up to about 28 s) is interpreted as the peak of the extended afterglow, whose temporal variability is produced by the interaction with the CBM. The numerical simulation that best reproduces the light curve (Fig.~\ref{fig:LC08}) and the time-integrated spectrum (Fig.~\ref{fig:spectrum08}) of this first episode ($3 s \leq t_a^d \leq 28 s$) is obtained with the following parameters: $E_{tot}^{e^\pm}=1.32\times10^{54}$ erg, $B=2.3 \times 10^{-3}$ and $\alpha=-1.8$; the Lorentz gamma factor at the transparency point, occurring at $r_0=2.8\times10^{14}$ cm, is $\gamma_0=428$. We consider an average number density $\left\langle n_{cbm}\right\rangle \sim 6$ particles cm$^{-3}$ and ${\cal R}=3.5\times10^{-10}$. The structure of the CBM adopted is presented in Fig.~\ref{fig:08density} and the adopted density contrast with respect to the average density is reported in Tab.~\ref{tab:08density}. The distribution of the CBM is just an approximation of the real one, where the CBM density shows some smooth fluctuations around its trend during the fireshell evolution. Nevertheless, it is sufficient to account for the observed variability in the luminosity. We must note that there is a sharp and short spike in the light curve $12.4$ s after the BAT trigger time \citep[see e.g.][]{2009AIPC.1133..356S} which we are unable to reproduce within our model, based on a spherically symmetric approximate dynamics. For this spike to be interpreted, a fully three-dimensional description of the CBM is needed. However, we expect that this more detailed description will not modify the overall dynamics of the system. In fact, the fluence of this spike is $\sim 4.9\%$ of the fluence observed in the first episode (between $2$ s and $28$ s) and $\sim 2.5\%$ of the fluence of the entire prompt emission (between $2$ s and $57$ s). Therefore, the error introduced by the omission of this spike from the numerical simulation is much smaller than the difference between the co-moving pure thermal spectrum and the modified one discussed in this paper. We can then omit the spike from our analysis, without hampering qualitatively the conclusions of our paper.

\begin{table}
\centering
       \begin{tabular}{|c|c|c|c|c|c|}
        \hline
& & & & &\\[-2.0ex]
peak &$r$ (cm)&$\Delta r$ (cm) & $\delta n/n$ & $M_{cloud} (M_{\odot})$ & $A_{vis} (cm)$ \\
& & & & &\\[-2.0ex]
\hline
& & & & &\\[-2.0ex]
a & $0.0$ & $6.0\times 10^{15}$ &$9.37$ & $5.7\times 10^{-9}$  & - \\
& & & & &\\[-2.0ex]
b &$6.0\times 10^{15}$ &$5.0 \times 10^{15}$ & $1.80$ & $6.3 \times 10^{-10}$ & $3.1 \times 10^{13}$\\
& & & & &\\[-2.0ex]
c &$1.1\times 10^{16}$ &$1.2\times 10^{16}$ & $0.55$ &$2.7 \times 10^{-9}$ & $5.7 \times 10^{13}$\\
& & & & &\\[-2.0ex]
d & $2.6\times 10^{16}$&$6.5\times 10^{15}$ & $0.47$ & $3.6 \times 10^{-10}$ & $1.4 \times 10^{14}$\\
& & & & &\\[-2.0ex]
e &$3.8\times 10^{16}$ &$1.5\times 10^{16}$ & $0.12$ & $1.1 \times 10^{-9}$& $2.2 \times 10^{14}$\\
& & & & &\\[-2.0ex]
f &$5.7\times 10^{16}$ & $3.0\times 10^{15}$& $0.33$ &$2.5 \times 10^{-11}$ & $3.5 \times 10^{14}$\\
& & & & &\\[-2.0ex]
g & $7.0\times 10^{16}$&$3.0\times 10^{16}$ & $0.14$ & $1.1 \times 10^{-11}$& $4.6 \times 10^{14}$\\  [0.5ex]
& & & & &\\[-2.0ex]
h & $7.6\times 10^{16}$&$3.0\times 10^{16}$ & $0.11$ & $8.3 \times 10^{-12}$& $5.2 \times 10^{14}$\\  [0.5ex]
        \hline
        \end{tabular}
        \caption{Properties of the CBM structure adopted for GRB 080319B: distance from the center of the explosion, thickness, normalized density and mass of the clumps; for each distance the transverse dimension of the visible area is also reported.}\label{tab:08density}
\end{table}

\begin{figure}
\centering
\includegraphics[width=\hsize,clip]{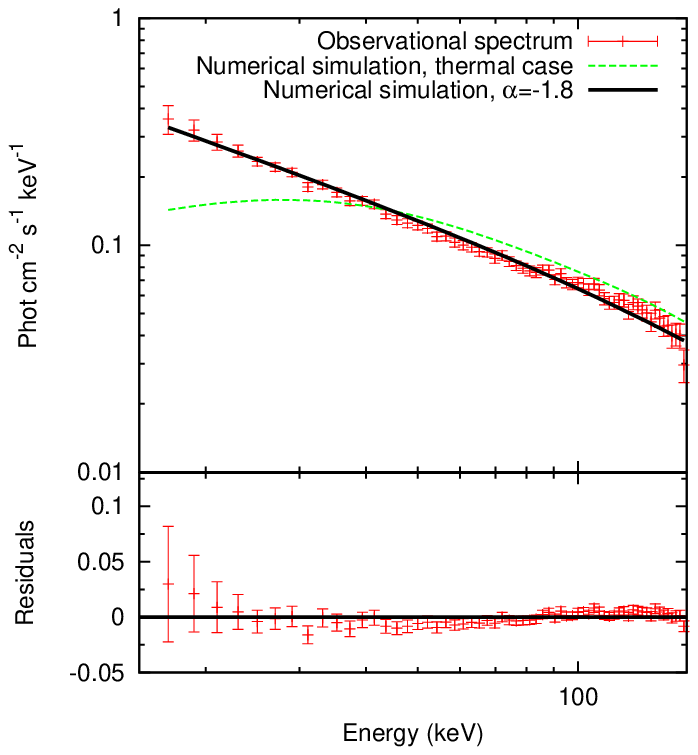}
\caption{Theoretically simulated spectra of GRB 080319B integrated over the time interval $3 s \leq t_a^d \leq 13 s$ with $\alpha=-1.8$ (black solid line) and $\alpha=0.0$ (pure thermal case, green dashed line) are compared with the data observed by BAT (red points). It can be seen that with the ``modified'' thermal spectrum we can correctly reproduce the observed spectrum, contrary to what happens with the pure thermal spectrum. In the residual plot the pure thermal case is omitted.}\label{fig:spectrum10s}
\end{figure}

With the above described set of parameters it is possible to interpret also successfully the spectra integrated over smaller intervals of time. Fig~\ref{fig:spectrum10s} shows, as an example, the spectrum for $3s \leq t_a^d \leq 13 s$: it can be seen that with the modified thermal spectrum we can correctly reproduce also this spectrum; on the contrary, by assuming a comoving thermal spectrum there are several discrepancies between the theoretical prediction and the observational data, especially at the lower energies. This is an important check to be made each time. In fact, changing the spectrum integration time means changing the number of different co-moving spectra which are convolved to get the observed one. The fact that the model is able to reproduce the observed spectrum regardless of the time scale over which it is integrated is therefore a clear support of the correctness of the assumed co-moving spectral shape.

To have an estimate of the sensitivity of the determination of these parameters of the model, we can proceed as follows. We fix $E_{tot}^{e^\pm}$ to the observed value of $1.32\times10^{54}$ erg \citep{2008GCN..7482....1G}. From the observational data we have that the fluence of the first 7 seconds, which correspond to the fluence of the P-GRB, $f_{P\hbox{-}GRB}$, is $2.19 \times 10^{-6} \leq f_{P\hbox{-}GRB} \lesssim 2.29 \times 10^{-6}$ erg/cm$^2$. This fixes a range of values for $B$: $2.19 \times 10^{-3} \lesssim B \leq 2.33 \times 10^{-3}$. Correspondingly, we must have $1.5 \times 10^{-10} \leq {\cal R} \leq 6.0 \times 10^{-10}$ and $4.1 \leq \langle n_{cbm} \rangle \leq 8.2$ particles/cm$^3$ to reproduce the observed light curves and spectra. It must be noted that the upper limit on $f_{P\hbox{-}GRB}$, and therefore the lower limit on $B$, is less stringent since we cannot exclude that more energy has been emitted in the P-GRB outside of the instrumental bandpass.

Concerning the second episode, lasting from 28 s to the end of the prompt emission, we performed numerical simulations with different sets of parameters, but we encountered several difficulties. In particular, while we can obtain a theoretical spectrum compatible with the observed one, it is not possible to correctly reproduce the time variability of the light curve, even when a bi-dimensional model for the CBM is adopted \citep{2006NCimB.121.1441B,2009AIPC.1111..383B}. This is consistent with the results presented by other authors: the time-resolved prompt emission spectra are best fit with power-laws and no change in the photon index is observed between the first and the second component \citep{2009AIPC.1133..356S}; on the contrary, a variation of the time-variability is found \citep{2008AIPC.1065..259M}. A possible explanation for this problem is that a fully three-dimensional modeling of the CBM is needed.

\section{GRB 050904}\label{GRB050904}

\begin{figure}
\centering
\includegraphics[width=\hsize,clip]{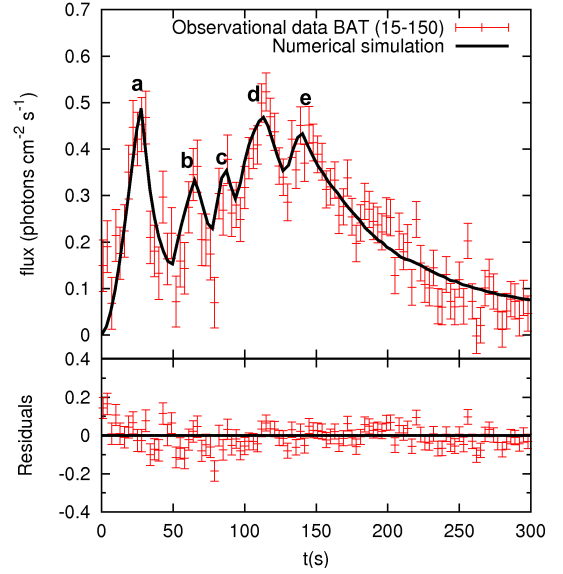}
\caption{Theoretically simulated light curve of GRB 050904 prompt emission in the 15-150 keV energy band (black solid curve) is compared with data observed by BAT (red points). The labels ``a'', ``b'',``c'',``d'' and ``e'' identify the peaks (see Fig.~\ref{fig:05density} and Tab.~\ref{tab:05density}).}\label{fig:LC05}
\end{figure}

\begin{figure}
\centering
\includegraphics[width=\hsize,clip]{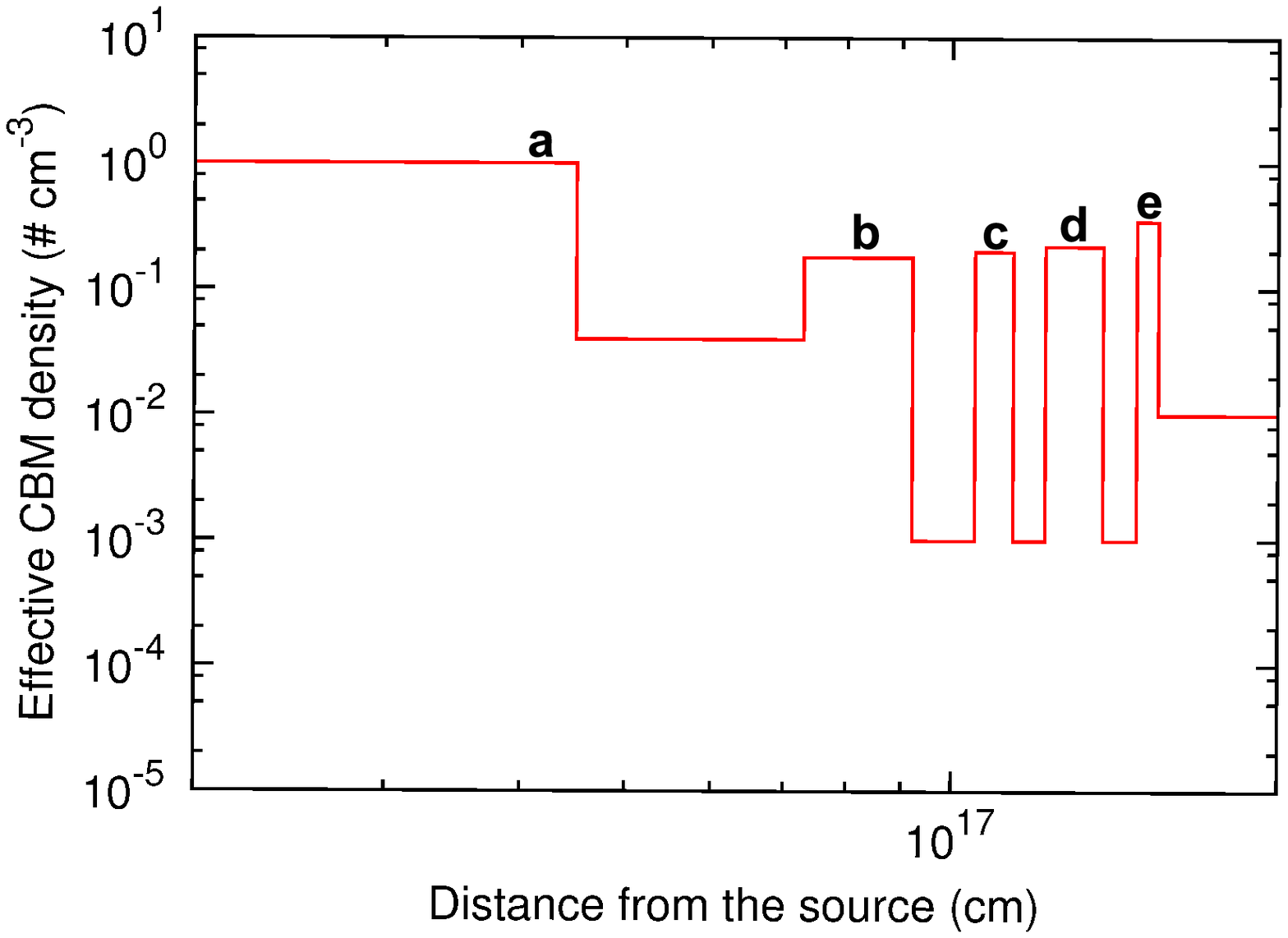}
\caption{Structure of the CBM adopted for GRB 050904. The labels ``a'', ``b'',``c'',``d'' and ``e'' indicate the values corresponding to the peaks in the BAT light curve (see Fig.~\ref{fig:LC05} and Tab.~\ref{tab:05density}).}\label{fig:05density}
\end{figure}

\begin{table}
\centering
       \begin{tabular}{|c|c|c|c|c|c|}
        \hline
& & & & &\\[-2.0ex]
peak &$r$ (cm)&$\Delta r$ (cm) & $\delta n/n$ & $M_{cloud} (M_{\odot})$ & $A_{vis} (cm)$ \\& & & & &\\[-2.0ex]
\hline
& & & & &\\[-2.0ex]
a &$0.0$ & $4.5\times 10^{16}$ & 4.3 & $4.0 \times 10^{-8}$& - \\
& & & & &\\[-2.0ex]
b & $7.3\times 10^{16}$& $1.9\times 10^{16}$& 0.8 & $5.4 \times 10^{-10}$ & $3.7 \times 10^{14}$ \\
& & & & &\\[-2.0ex]
c & $1.05\times 10^{17}$&$9.0\times 10^{15}$ &0.9 & $6.4 \times 10^{-11}$&  $5.7 \times 10^{14}$\\
& & & & &\\[-2.0ex]
d & $1.22\times 10^{17}$& $1.6\times 10^{16}$& 1.0 & $4.0 \times 10^{-10}$& $7.1 \times 10^{14}$\\
& & & & &\\[-2.0ex]
e & $1.48\times 10^{17}$& $7.0\times 10^{15}$& 1.5 & $5.3 \times 10^{-11}$& $1.0 \times 10^{15}$\\  [0.5ex]
        \hline
        \end{tabular}
        \caption{Properties of the CBM structure adopted for GRB 050904: distance from the center of the explosion, thickness, normalized density and mass of the clumps; for each distance the transverse dimension of the visible area is also reported.}\label{tab:05density}
\end{table}

\begin{figure}
\centering
\includegraphics[width=\hsize,clip]{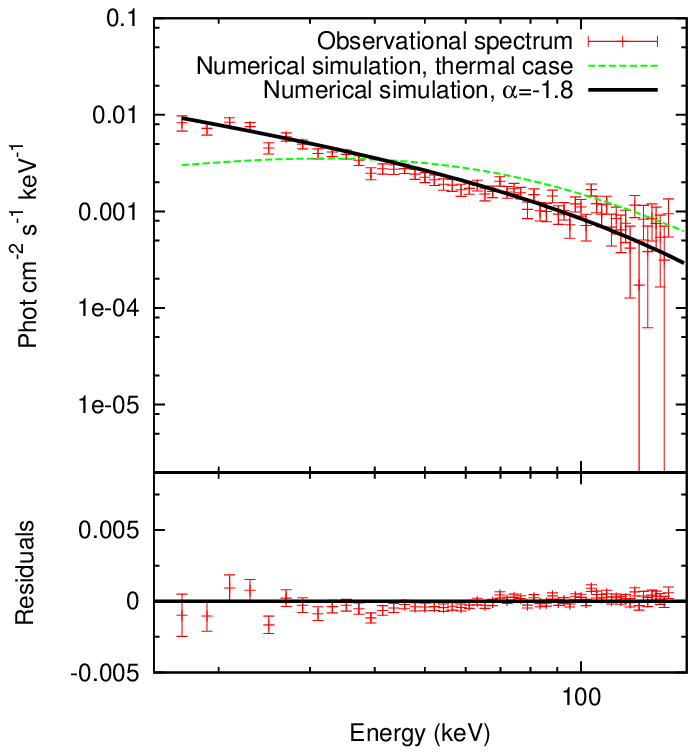}
\caption{Theoretically simulated time integrated spectra of GRB 050904 for $0\leq t_a^d \leq 225 s$ with $\alpha=-1.8$ (black solid line) and $\alpha=0.0$ (pure thermal case, green dashed line) are compared with the data observed by BAT (red points). It can be seen that with the ``modified'' thermal spectrum we can correctly reproduce the observed spectrum, contrary to what happens with the pure thermal spectrum. In the residual plot the pure thermal case is omitted. The range of vertical axes in the residual plot has been chosen to be the same of Fig.~\ref{fig:spectrum0550s}.}\label{fig:spectrum05}
\end{figure}

We analyzed the prompt emission light curve (Fig.~\ref{fig:LC05}) and spectrum (Fig.~\ref{fig:spectrum05}) of GRB 050904 observed by BAT. As we already mentioned in the Introduction, for GRB 050904 we have $E_{peak}^{obs}=314^{+173}_{-89}$ keV \citep{2009PASJ...61..521S}. The data have been obtained by using the standard FTOOLS package (Heasoft, version 6.10); the BAT spectrum integrated over the $T_{90}$ of the source ($T_{90}=225 \pm 10$ s, see \citealp{2005GCN..3938....1S}) is best modeled with a power-law with photon index $\gamma$=1.25 $\pm$ 0.07, with a chi square value of $\chi^2=64.09$ for 60 degrees of freedom. Within the fireshell model, we identify the prompt emission with the peak of the extended afterglow. In this case the P-GRB has not been observed. In fact, we have estimated $E^{iso}_{P\hbox{-}GRB}=1.99\times10^{52}$ erg, that for $z=6.29$ corresponds to a fluence of $\sim 6.3\times10^{-9}$ erg cm$^{-2}$. If we assume an observed duration $\Delta t_{P\hbox{-}GRB}\gtrsim 1$ s, the P-GRB flux is under the BAT threshold. The numerical simulation that best reproduce the observational data is obtained with similar values of $E_{tot}^{e^\pm}$ and $B$ found for GRB 080319B: $E_{tot}^{e^\pm}=1.0\times10^{54}$ erg and $B=2.2\times 10^{-3}$, with a Lorentz gamma factor at the transparency point $\gamma_0=446$. This could be an indication of a similar progenitor for the two sources. Concerning the other model parameters, we found an average number density $\left\langle n_{cbm}\right\rangle \sim 0.2$ particles cm$^{-3}$ and ${\cal R}=2\times10^{-11}$; these values are different from the ones obtained for GRB 080319B and this could be an indication of the fact that the two bursts occurred in different environments. The structure of the CBM adopted is shown in Fig.~\ref{fig:05density} and the adopted density contrast with respect to the mean density is reported in Tab.~\ref{tab:05density}. 

\begin{figure}
\centering
\includegraphics[width=\hsize,clip]{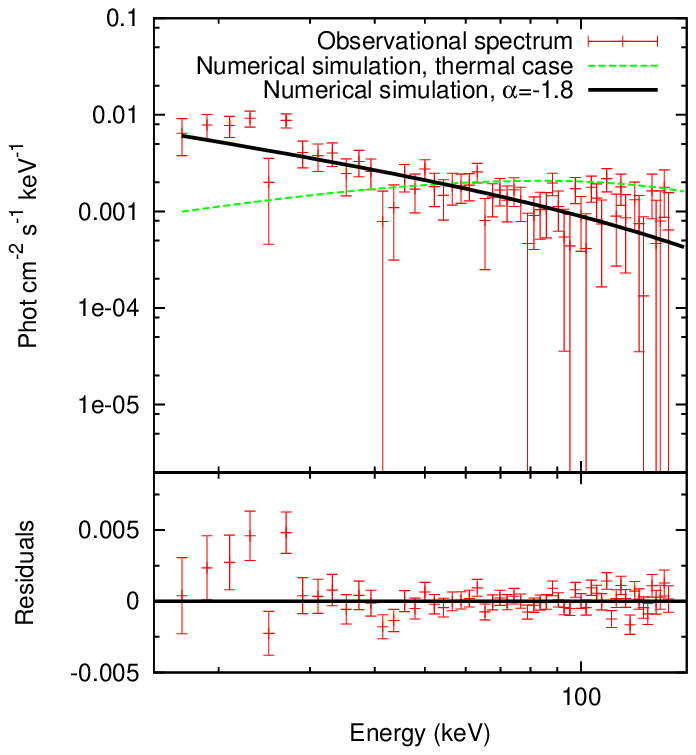}
\caption{Theoretically simulated time integrated spectra of GRB 050904 for $0\leq t_a^d \leq 50 s$ with $\alpha=-1.8$ (black solid line) and $\alpha=0.0$ (pure thermal case, green dashed line) are compared with the data observed by BAT (red points). It can be seen that with the ``modified'' thermal spectrum we can correctly reproduce the observed spectrum, contrary to what happens with the pure thermal spectrum. In the residual plot the pure thermal case is omitted.}\label{fig:spectrum0550s}
\end{figure}

Also in this case the numerical simulation that best reproduces the observational data has been obtained assuming the value $-1.8$ for the free parameter $\alpha$; in this way we can also correctly reproduce spectra integrated over intervals of time much less than the $T_{90}$ of the source (in Fig.~\ref{fig:spectrum0550s} is shown, as an example, the BAT spectrum integrated over the first 50 s). Once again, the fact that the model is able to reproduce the observed spectrum regardless of the time scale over which it is integrated is a clear support of the correctness of the assumed co-moving spectral shape.

Estimating the sensitivity of the determination of these parameters of the model in this case of GRB 050904 is more difficult than in the previous case of GRB 080319B. In fact, in this case the P-GRB is not observed and therefore we may have only a lower limit on the value of $B$. However, we can proceed as follows. We fix $E_{tot}^{e^\pm}$ to the observed value of $E_{iso}=1.04^{+0.25}_{-0.17}\times 10^{54}$ erg \citep{2009PASJ...61..521S}, i.e. $8.7 \times 10^{53} \leq E_{tot}^{e^\pm} \leq 1.29 \times 10^{54}$ erg. We can then make an educated guess about the average value of $n_{cbm}$, i.e. we can assume that $0.1 \lesssim \langle n_{cbm} \rangle \lesssim 10$ particles/cm$^3$ (see Sec.~\ref{procedure}). With this choice, we obtain $1.9 \times 10^{-3} \lesssim B \lesssim 3.4 \times 10^{-3}$ and $1.5 \times 10^{-11} \lesssim {\cal R} \lesssim 8.0 \times 10^{-11}$ to reproduce the observed light curves and spectra.

\section{Discussion on the comoving spectrum}\label{sec:disc}

We have mentioned in Sec.~\ref{intro} that GRBs with $E_{iso}$ up to $10^{53}$ erg have been successfully interpreted within the traditional fireshell model by assuming a pure comoving thermal spectrum. In the previous sections we have also shown that a modification of the comoving spectrum (see Eq.~\ref{eq:denfmod}) is needed to correctly reproduce the observational data of two of the most energetic GRBs, GRB 080319B and GRB 050904, within the fireshell model: the difficulty in interpreting the BAT data by assuming a pure comoving thermal spectrum for these sources has been clearly shown in Fig.~\ref{fig:spectrum10s}.

The reasons of this result can be summarized as follows (see also Sec.~\ref{intro}):

\begin{enumerate}
\item The modification of the spectral energy distribution given in Eq.~\ref{eq:denfmod} does not affect the spectrum near the peak, but it affects the low energy Rayleigh-Jeans tail of the distribution (see Figs.~\ref{fig:sp20s} and \ref{fig:spectra}).
\item For the sources with $E_{iso} \lesssim 10^{53}$ erg previously analysed within the fireshell model, only the spectral region around the peak contributes to the emission in the instrument bandpasses. We give two explicit examples in Tab.~\ref{tab:epeak}: the INTEGRAL and \emph{Swift} BAT observations of GRB 031203 and GRB 060607A only cover the region around the peak.
\item In the case of the most energetic sources, like GRB 080319B and GRB 050904, $E_{peak}^{obs}$ is outside the \emph{Swift} BAT bandpass (see Tab.~\ref{tab:epeak}). Conversely, the \emph{Swift} BAT data cover precisely the low energy component of the spectrum, where the effect of the new parametrization of the spectrum is maximised (see Tab.~\ref{tab:epeak}).
\end{enumerate}

\begin{table}
\centering
{\scriptsize       
       \begin{tabular}{|c|c|c|c|c|}
        \hline 
GRB & 031203 (a)& 060607A (b)& 080319B (c)& 050904 (d)\\
\hline
& & & & \\[-2.0ex]
z & 0.106 &3.082 &0.937 &6.29 \\[1.2ex]
$E_{iso}$ (erg) &$10^{50}$ & $1.1 \times 10^{53}$ &$1.34\times 10^{54}$ &$1.04 \times 10^{54}$ \\[1ex]
$E_{peak}^{i}$ (keV) & $158 \pm 51$ &$478^{+314}_{-69}$ &1261 $\pm$ 65 & $2291^{+1263}_{-634}$\\[1.2ex]
$E_{peak}^{obs}$ (keV) &$144 \pm 46$ & $117^{+77}_{-17}$& 675 $\pm$ 22& $314^{+173}_{-89}$\\[1.2ex]
Satellite&INTEGRAL&\emph{Swift}&\emph{Swift}&\emph{Swift}\\[1.2ex]
Instrument&IBIS/ISGRI&BAT&BAT&BAT \\[1.2ex]
$\Delta E$ (keV) &17 -- 500 & 15 -- 150& 15 -- 150&15 -- 150 \\
\hline
        \end{tabular}        
}        
        \caption{Redshift ($z$), isotropic energy ($E_{iso}$), intrinsic and observed peak energy ($E_{peak}^{i}$ and $E_{peak}^{obs}$ respectively) of GRB spectra analysed within the fireshell model. The instruments by which the observational data interpreted within the fireshell model have been taken, together with the energy range they cover ($\Delta E$), are also reported. 
References for z: (a) \citet{2004ApJ...611..200P}; (b) \citet{2006GCN..5237....1L}; (c) \citet{2008GCN..7444....1V}; (d) \citet{2005GCN..3937....1K}. References for $E_{iso}$: (a) \citet{2006MNRAS.372..233A}; (b) L. Amati, private communication; (c) \citet{2008GCN..7482....1G} ; (d) \citet{2009PASJ...61..521S}. References for $E_{peak}^i$: (a) \citet{2006MNRAS.372..233A}; (b) L. Amati, private communication; (c) \citet{2008GCN..7482....1G} ; (d) \citet{2009PASJ...61..521S}. References for $E_{peak}^{obs}$: (a) \citet{2005NCimC..28..351U} ; (b) L. Amati, private communication; (c) \citet{2008Natur.455..183R}; (d) \citet{2009PASJ...61..521S}.}\label{tab:epeak}
\end{table}

The issue if the modification of the comoving spectrum given in Eq.~\ref{eq:denfmod} is really universal and applies as well to lower energetic sources is still open. It could only be settled by future space missions dedicated to the observation of the prompt emission below 10 keV such as LOFT \citep{2011arXiv1107.0436F} and MIRAX \citep{mirax}.

As correctly pointed out to us by an anonymous referee, a critical test for the modified comoving thermal spectrum can also come from GRB 061121 \citep{2007ApJ...663.1125P}. In this special case, a soft precursor pulse triggered BAT, which allowed Swift to slew in time for BAT, XRT \citep{2006GCN..5832....1P} and UVOT to simultaneously observe the prompt emission. Furthermore, Konus-Wind \citep{2006GCN..5837....1G} also observed this burst, which resulted in an $E_{peak}^{obs} \sim 606 +90/-72$ keV (well above the BAT bandpass) and $E_{iso} \sim 2.5\times 10^{53}$ erg, given a redshift of $z = 1.314$ \citep{2006GCN..5826....1B}. This source presents many other interesting and challenging observational features, and a complete analysis is going to be presented in a separate paper. However, we like to emphasize that the photon index of the XRT data during the main event of the prompt emission is $0.6 \lesssim \Gamma \lesssim 0.8$ \footnote{see e.g. data at \url{http://www.swift.ac.uk/burst\_analyser/00239899/}.}, in very good agreement with the expected value from the modified thermal spectrum assuming $\alpha=-1.8$, which is indeed $\sim 0.8$.

\section{Conclusions}\label{concl}

GRB 050904 was discovered in the pre-\emph{Fermi} and pre-AGILE era, while GRB 080319B was discovered in the pre-\emph{Fermi} era and was unobservable by AGILE due to Earth occultation. With the exception of the data from \emph{Konus}-WIND and \emph{Suzaku}-WAM, no observations on the high energy component of these two sources are available. The high quality data from the BAT instrument on board the \emph{Swift} satellite, nevertheless, have allowed us to reach a quite firm conclusion on the low energy component of the spectra of these sources.

Thanks to these most energetic sources we have been able, for the first time, to explore the Rayleigh-Jeans tail of the comoving blackbody spectral energy distribution and to conclude that it must be modified with an additional component. We recall, in fact, that in the original proposal the thermal nature of the spectrum in the comoving frame was adopted only for simplicity, inspired by a similar approach followed by Enrico Fermi in the different context of ultra high-energy collisions. Notice that even photospheric emission in ultrarelativistically expanding sources does not produce pure thermal spectra (\citealp{2011arXiv1110.0407R}; see also \citealp{2011ApJ...732...49P}). The successful interpretation of many sources \citep[see e.g.][]{2005ApJ...634L..29B,2006ApJ...645L.109R,2007A&A...471L..29D,2007A&A...474L..13B,2009A&A...498..501C,2010A&A...521A..80C,2011A&A...529A.130D} showed the viability of this ansatz \citep{2004IJMPD..13..843R}. In the intervening years, thanks to the data analysis by Felix Ryde and collaborators \citep[see e.g.][]{2004ApJ...614..827R,2009ApJ...702.1211R}, it has become clear that the existence of a pure black body spectra in any GRB observation is more an exception than the rule (see Sec.~\ref{phen}). In the present work we show that this is also the case for the comoving spectrum of the extended afterglow. The most interesting aspect is that it is possible to generalize the previous ansatz by the addition of a single power-law component in order to recover a consistent interpretation of all previous results and of the ones corresponding to the present more energetic sources. We have introduced a new phenomenological parameter $\alpha$ describing such an additional component. The choice of $\alpha=-1.8$ leads to a coherent description of both sources, not contradictory with the previous results on the less energetic sources. The main goal of our work in this paper is to maximize the knowledge acquirable for these very energetic sources ($E_{iso} \sim 10^{54}$ erg), which have a peak of emission at energies much higher than the less energetic ones ($E_{p,i} > 1$ MeV), and therefore to explore the low energy part of the spectrum of the prompt emission. A strong and promising theoretical activity is currently devoted to ascertain a possible role of collisionless shocks in generating power law components in the high energy part of
the photon spectrum \citep{2008ApJ...682L...5S}. Also, the synchrotron ``line of death'' appears to be problematic for such models \citep{2009ApJ...707L..92S}. The knowledge of the spectrum in the comoving frame is certainly an important step toward the identification of the physical process occurring in the interaction of the accelerated baryons with the CBM, which is yet largely unknown.

As an additional result, the analysis of GRB light curves and spectra within the fireshell model allows us to infer the filamentary, clumpy and porosity structure of the CBM. Specifically, we determined $\langle n_{cbm} \rangle= 6$ particles cm$^{-3}$ and ${\cal R}=3.5\times 10^{-10}$ for GRB 080319B, $\langle n_{cbm} \rangle$= 0.2 particles cm$^{-3}$ and ${\cal R}=2.0\times 10^{-11}$ for GRB 050904 (see also Tab.~\ref{FinTab}).

\begin{table}
\centering
\begin{tabular}{|c|c|c|}
\hline
GRB & GRB 080319B & GRB 050904\\
\hline
$E_{tot}^{e^\pm}$ (erg)& $1.32\times10^{54}$ & $1.0\times10^{54}$ \\
$B$ & $2.3 \times 10^{-3}$ & $2.2\times 10^{-3}$\\
$\alpha$ & $-1.8$ & $-1.8$\\
$\left\langle n_{cbm}\right\rangle$ (\#/cm$^3$)& $\sim 6$ & $\sim 0.2$\\
${\cal R}$ & $3.5\times10^{-10}$ & $2\times10^{-11}$\\
\hline
\end{tabular}
\caption{Summary of parameters characterizing GRB 080319B and GRB 050904.}\label{FinTab}
\end{table}

We can correctly reproduce the whole BAT prompt emission data of GRB 050904. For GRB 080319B only the first $\sim 28$ s have been satisfactorily interpreted within the fireshell model with a mono-dimensional CBM description. Concerning the remaining part of the prompt emission, it occurs at a distance $r > 10^{17}$ cm, where the transverse dimension of the visible area is much larger than the typical size of the CBM clumps ($5.0\times 10^{15} \lesssim \Delta r \lesssim 1.7 \times 10^{16}$ cm, see Tab.~\ref{tab:08density}). Therefore, a fully three-dimensional modeling of the CBM is needed at such a distance.

We have mentioned in Sec.~\ref{sec:SEDprompt} the central role of the phenomenological parameterizations in the description of GRBs. In addition to the $\alpha$ parameter here defined, we have recalled the \citet{1993ApJ...413..281B} formula, the \citet{2002A&A...390...81A} relation and the coefficients described by \citet{2004ApJ...614..827R,2005ApJ...625L..95R} and \citet{2009ApJ...702.1211R}. Although no physical explanation for these parameters have been reached, they represent certainly a fundamental step in reaching a quantitative and qualitative description of the source and help to understand the underlying physical process of GRBs.

In the fireshell model the radiation observed in the BAT data comes from an integration which takes into account the CBM filamentary structure and applies a double convolution, over the EQTS and the observation time, of a mixing of the co-moving thermal and power-law components given by Eq.(\ref{eq:denfmod}). In the BAT energy range the cutoffs of the thermal components give a fundamental contribution. It was therefore unexpected that the theoretically computed spectrum would have given rise to a power-law so closely resembling the observed one. There is no simple relation between the power-law index of the observed BAT spectrum and the one of the power-law component in the co-moving spectrum given in Eq.(\ref{eq:denfmod}). When the co-moving thermal component becomes negligible at low enough energy, the co-moving spectrum is described by just the power-law component. The convolution of power-laws with the same index results in a power-law of that index. Observations by XRT in the prompt emission of a highly energetic source, if available, may then give direct and independent information about the existence of the power-law component in Eq.(\ref{eq:denfmod}) and on its index. If so confirmed, this power-law component would not be just a mere phenomenological optimization of the agreement between our theory and the observed BAT spectra. It would be an independent physical component of the co-moving spectrum, whose index can be directly read from the observational data. This would give an additional strong confirmation of our model.

\acknowledgements
We thank C. Guidorzi for the reduced \emph{Swift} BAT data of GRB 080319B. We are especially grateful to an anonymous referee for her/his important remarks which have improved the presentation of our results.


\begin{thebibliography}{156}
\expandafter\ifx\csname natexlab\endcsname\relax\def\natexlab#1{#1}\fi

\bibitem[{{Abdo} {et~al.}(2009){Abdo}, {Ackermann}, {Ajello}, {Asano},
  {Atwood}, {Axelsson}, {Baldini}, {Ballet}, {Barbiellini}, {Baring},
  {Bastieri}, {Bechtol}, {Bellazzini}, {Berenji}, {Bhat}, {Bissaldi},
  {Blandford}, {Bloom}, {Bonamente}, {Borgland}, {Bouvier}, {Bregeon}, {Brez},
  {Briggs}, {Brigida}, {Bruel}, {Burgess}, {Burrows}, {Buson}, {Caliandro},
  {Cameron}, {Caraveo}, {Casandjian}, {Cecchi}, {{\c C}elik}, {Chekhtman},
  {Cheung}, {Chiang}, {Ciprini}, {Claus}, {Cohen-Tanugi}, {Cominsky},
  {Connaughton}, {Conrad}, {Cutini}, {d'Elia}, {Dermer}, {de Angelis}, {de
  Palma}, {Digel}, {Dingus}, {Silva}, {Drell}, {Dubois}, {Dumora}, {Farnier},
  {Favuzzi}, {Fegan}, {Finke}, {Fishman}, {Focke}, {Fortin}, {Frailis},
  {Fukazawa}, {Funk}, {Fusco}, {Gargano}, {Gehrels}, {Germani}, {Giavitto},
  {Giebels}, {Giglietto}, {Giordano}, {Glanzman}, {Godfrey}, {Goldstein},
  {Granot}, {Greiner}, {Grenier}, {Grove}, {Guillemot}, {Guiriec}, {Hanabata},
  {Harding}, {Hayashida}, {Hays}, {Horan}, {Hughes}, {Jackson},
  {J{\'o}hannesson}, {Johnson}, {Johnson}, {Johnson}, {Kamae}, {Katagiri},
  {Kataoka}, {Kawai}, {Kerr}, {Kippen}, {Kn{\"o}dlseder}, {Kocevski}, {Komin},
  {Kouveliotou}, {Kuss}, {Lande}, {Latronico}, {Lemoine-Goumard}, {Longo},
  {Loparco}, {Lott}, {Lovellette}, {Lubrano}, {Madejski}, {Makeev},
  {Mazziotta}, {McBreen}, {McEnery}, {McGlynn}, {Meegan}, {M{\'e}sz{\'a}ros},
  {Meurer}, {Michelson}, {Mitthumsiri}, {Mizuno}, {Moiseev}, {Monte},
  {Monzani}, {Moretti}, {Morselli}, {Moskalenko}, {Murgia}, {Nakamori},
  {Nolan}, {Norris}, {Nuss}, {Ohno}, {Ohsugi}, {Omodei}, {Orlando}, {Ormes},
  {Paciesas}, {Paneque}, {Panetta}, {Pelassa}, {Pepe}, {Pesce-Rollins},
  {Petrosian}, {Piron}, {Porter}, {Preece}, {Rain{\`o}}, {Rando}, {Rau},
  {Razzano}, {Razzaque}, {Reimer}, {Reimer}, {Reposeur}, {Ritz}, {Rochester},
  {Rodriguez}, {Roming}, {Roth}, {Ryde}, {Sadrozinski}, {Sanchez}, {Sander},
  {Saz Parkinson}, {Scargle}, {Schalk}, {Sgr{\`o}}, {Siskind}, {Smith},
  {Spinelli}, {Stamatikos}, {Stecker}, {Stratta}, {Strickman}, {Suson},
  {Swenson}, {Tajima}, {Takahashi}, {Tanaka}, {Thayer}, {Thayer}, {Thompson},
  {Tibaldo}, {Torres}, {Tosti}, {Tramacere}, {Uchiyama}, {Uehara}, {Usher},
  {van der Horst}, {Vasileiou}, {Vilchez}, {Vitale}, {von Kienlin}, {Waite},
  {Wang}, {Wilson-Hodge}, {Winer}, {Wood}, {Yamazaki}, {Ylinen}, \&
  {Ziegler}}]{2009ApJ...706L.138A}
{Abdo}, A.~A., {Ackermann}, M., {Ajello}, M., {et~al.} 2009, ApJ, 706, L138

\bibitem[{{Ackermann} {et~al.}(2010){Ackermann}, {Asano}, {Atwood}, {Axelsson},
  {Baldini}, {Ballet}, {Barbiellini}, {Baring}, {Bastieri}, {Bechtol},
  {Bellazzini}, {Berenji}, {Bhat}, {Bissaldi}, {Blandford}, {Bloom},
  {Bonamente}, {Borgland}, {Bouvier}, {Bregeon}, {Brez}, {Briggs}, {Brigida},
  {Bruel}, {Buson}, {Caliandro}, {Cameron}, {Caraveo}, {Carrigan},
  {Casandjian}, {Cecchi}, {{\c C}elik}, {Charles}, {Chiang}, {Ciprini},
  {Claus}, {Cohen-Tanugi}, {Connaughton}, {Conrad}, {Dermer}, {de Palma},
  {Dingus}, {Silva}, {Drell}, {Dubois}, {Dumora}, {Farnier}, {Favuzzi},
  {Fegan}, {Finke}, {Focke}, {Frailis}, {Fukazawa}, {Fusco}, {Gargano},
  {Gasparrini}, {Gehrels}, {Germani}, {Giglietto}, {Giordano}, {Glanzman},
  {Godfrey}, {Granot}, {Grenier}, {Grondin}, {Grove}, {Guiriec}, {Hadasch},
  {Harding}, {Hays}, {Horan}, {Hughes}, {J{\'o}hannesson}, {Johnson}, {Kamae},
  {Katagiri}, {Kataoka}, {Kawai}, {Kippen}, {Kn{\"o}dlseder}, {Kocevski},
  {Kouveliotou}, {Kuss}, {Lande}, {Latronico}, {Lemoine-Goumard}, {Llena
  Garde}, {Longo}, {Loparco}, {Lott}, {Lovellette}, {Lubrano}, {Makeev},
  {Mazziotta}, {McEnery}, {McGlynn}, {Meegan}, {M{\'e}sz{\'a}ros}, {Michelson},
  {Mitthumsiri}, {Mizuno}, {Moiseev}, {Monte}, {Monzani}, {Moretti},
  {Morselli}, {Moskalenko}, {Murgia}, {Nakajima}, {Nakamori}, {Nolan},
  {Norris}, {Nuss}, {Ohno}, {Ohsugi}, {Omodei}, {Orlando}, {Ormes}, {Ozaki},
  {Paciesas}, {Paneque}, {Panetta}, {Parent}, {Pelassa}, {Pepe},
  {Pesce-Rollins}, {Piron}, {Preece}, {Rain{\`o}}, {Rando}, {Razzano},
  {Razzaque}, {Reimer}, {Ritz}, {Rodriguez}, {Roth}, {Ryde}, {Sadrozinski},
  {Sander}, {Scargle}, {Schalk}, {Sgr{\`o}}, {Siskind}, {Smith}, {Spandre},
  {Spinelli}, {Stamatikos}, {Stecker}, {Strickman}, {Suson}, {Tajima},
  {Takahashi}, {Takahashi}, {Tanaka}, {Thayer}, {Thayer}, {Thompson},
  {Tibaldo}, {Toma}, {Torres}, {Tosti}, {Tramacere}, {Uchiyama}, {Uehara},
  {Usher}, {van der Horst}, {Vasileiou}, {Vilchez}, {Vitale}, {von Kienlin},
  {Waite}, {Wang}, {Wilson-Hodge}, {Winer}, {Wu}, {Yamazaki}, {Yang}, {Ylinen},
  \& {Ziegler}}]{2010ApJ...716.1178A}
{Ackermann}, M., {Asano}, K., {Atwood}, W.~B., {et~al.} 2010, ApJ, 716, 1178

\bibitem[{{Ackermann} {et~al.}(2011){Ackermann}, {Ajello}, {Asano}, {Axelsson},
  {Baldini}, {Ballet}, {Barbiellini}, {Baring}, {Bastieri}, {Bechtol},
  {Bellazzini}, {Berenji}, {Bhat}, {Bissaldi}, {Blandford}, {Bonamente},
  {Borgland}, {Bouvier}, {Bregeon}, {Brez}, {Briggs}, {Brigida}, {Bruel},
  {Buehler}, {Buson}, {Caliandro}, {Cameron}, {Caraveo}, {Carrigan},
  {Casandjian}, {Cecchi}, {{\c C}elik}, {Chaplin}, {Charles}, {Chekhtman},
  {Chiang}, {Ciprini}, {Claus}, {Cohen-Tanugi}, {Connaughton}, {Conrad},
  {Cutini}, {Dermer}, {de Angelis}, {de Palma}, {Dingus}, {Silva}, {Drell},
  {Dubois}, {Favuzzi}, {Fegan}, {Ferrara}, {Focke}, {Frailis}, {Fukazawa},
  {Funk}, {Fusco}, {Gargano}, {Gasparrini}, {Gehrels}, {Germani}, {Giglietto},
  {Giordano}, {Giroletti}, {Glanzman}, {Godfrey}, {Goldstein}, {Granot},
  {Greiner}, {Grenier}, {Grove}, {Guiriec}, {Hadasch}, {Hanabata}, {Harding},
  {Hayashi}, {Hayashida}, {Hays}, {Horan}, {Hughes}, {Itoh}, {J{\'o}hannesson},
  {Johnson}, {Johnson}, {Kamae}, {Katagiri}, {Kataoka}, {Kippen},
  {Kn{\"o}dlseder}, {Kocevski}, {Kouveliotou}, {Kuss}, {Lande}, {Latronico},
  {Lee}, {Llena Garde}, {Longo}, {Loparco}, {Lovellette}, {Lubrano}, {Makeev},
  {Mazziotta}, {McBreen}, {McEnery}, {McGlynn}, {Meegan}, {Mehault},
  {M{\'e}sz{\'a}ros}, {Michelson}, {Mizuno}, {Monte}, {Monzani}, {Moretti},
  {Morselli}, {Moskalenko}, {Murgia}, {Nakajima}, {Nakamori}, {Naumann-Godo},
  {Nishino}, {Nolan}, {Norris}, {Nuss}, {Ohno}, {Ohsugi}, {Okumura}, {Omodei},
  {Orlando}, {Ormes}, {Ozaki}, {Paciesas}, {Paneque}, {Panetta}, {Parent},
  {Pelassa}, {Pepe}, {Pesce-Rollins}, {Petrosian}, {Piron}, {Porter}, {Preece},
  {Racusin}, {Rain{\`o}}, {Rando}, {Rau}, {Razzano}, {Razzaque}, {Reimer},
  {Reimer}, {Reposeur}, {Reyes}, {Ripken}, {Ritz}, {Roth}, {Ryde},
  {Sadrozinski}, {Sander}, {Scargle}, {Schalk}, {Sgr{\`o}}, {Siskind}, {Smith},
  {Spandre}, {Spinelli}, {Stamatikos}, {Stecker}, {Strickman}, {Suson},
  {Tajima}, {Takahashi}, {Tanaka}, {Tanaka}, {Thayer}, {Thayer}, {Tibaldo},
  {Tierney}, {Toma}, {Torres}, {Tosti}, {Tramacere}, {Uchiyama}, {Uehara},
  {Usher}, {Vandenbroucke}, {van der Horst}, {Vasileiou}, {Vilchez}, {Vitale},
  {von Kienlin}, {Waite}, {Wang}, {Wilson-Hodge}, {Winer}, {Wood}, {Wu},
  {Yamazaki}, {Yang}, {Ylinen}, \& {Ziegler}}]{2011ApJ...729..114A}
{Ackermann}, M., {Ajello}, M., {Asano}, K., {et~al.} 2011, ApJ, 729, 114

\bibitem[{{Aksenov} {et~al.}(2007){Aksenov}, {Ruffini}, \&
  {Vereshchagin}}]{2007PhRvL..99l5003A}
{Aksenov}, A., {Ruffini}, R., \& {Vereshchagin}, G. 2007, Phys. Rev. Lett., 99,
  125003

\bibitem[{{Aksenov} {et~al.}(2010){Aksenov}, {Bernardini}, {Bianco}, {Caito},
  {Cherubini}, {De Barros}, {Geralico}, {Izzo}, {Massucci}, {Patricelli},
  {Rotondo}, {Rueda Hernandez}, {Ruffini}, {Vereshchagin}, \&
  {Xue}}]{aksenovVenice}
{Aksenov}, A.~G., {Bernardini}, M.~G., {Bianco}, C.~L., {et~al.} 2010, in SIF
  Conference Proceedings, Vol. 102, The Shocking Universe, ed. G.~{Chincarini},
  P.~{D'Avanzo}, R.~{Margutti}, \& R.~{Salvaterra}, 451

\bibitem[{{Amati}(2006)}]{2006MNRAS.372..233A}
{Amati}, L. 2006, MNRAS, 372, 233

\bibitem[{{Amati} {et~al.}(2009){Amati}, {Frontera}, \&
  {Guidorzi}}]{2009A&A...508..173A}
{Amati}, L., {Frontera}, F., \& {Guidorzi}, C. 2009, A\&A, 508, 173

\bibitem[{{Amati} {et~al.}(2008){Amati}, {Guidorzi}, {Frontera}, {Della Valle},
  {Finelli}, {Landi}, \& {Montanari}}]{2008MNRAS.391..577A}
{Amati}, L., {Guidorzi}, C., {Frontera}, F., {et~al.} 2008, MNRAS, 391, 577

\bibitem[{{Amati} {et~al.}(2002){Amati}, {Frontera}, {Tavani}, {in't Zand},
  {Antonelli}, {Costa}, {Feroci}, {Guidorzi}, {Heise}, {Masetti}, {Montanari},
  {Nicastro}, {Palazzi}, {Pian}, {Piro}, \& {Soffitta}}]{2002A&A...390...81A}
{Amati}, L., {Frontera}, F., {Tavani}, M., {et~al.} 2002, A\&A, 390, 81

\bibitem[{{Amati} {et~al.}(2011){Amati}, {Feroci}, {Frontera}, {Labanti},
  {Vacchi}, {Argan}, {Campana}, {Costa}, {Ruffini}, {Bombaci}, {Del Monte},
  {Donnarumma}, {Drago}, {Evangelista}, {Farinelli}, {Ghirlanda}, {Ghisellini},
  {Guidorzi}, {Fuschino}, {Lazzarotto}, {Lazzati}, {Malcovati}, {Marisaldi},
  {Morelli}, {Muleri}, {Orlandini}, {Pacciani}, {Pian}, {Rapisarda}, {Rubini},
  {Salvaterra}, {Soffitta}, {Titarchuk}, {Traci}, {Rashevsky}, {Zampa},
  {Zampa}, {Auricchio}, {Basili}, {Caroli}, {Maiorano}, {Masetti}, {Nicastro},
  {Palazzi}, {Silvestri}, {Stephen}, \& {Braga}}]{mirax}
{Amati}, L., {Feroci}, M., {Frontera}, F., {et~al.} 2011, Nuovo Cimento C, 34,
  49

\bibitem[{{Aptekar} {et~al.}(1995){Aptekar}, {Frederiks}, {Golenetskii},
  {Ilynskii}, {Mazets}, {Panov}, {Sokolova}, {Terekhov}, {Sheshin}, {Cline}, \&
  {Stilwell}}]{1995SSRv...71..265A}
{Aptekar}, R.~L., {Frederiks}, D.~D., {Golenetskii}, S.~V., {et~al.} 1995,
  Space Science Reviews, 71, 265

\bibitem[{{Atwood} {et~al.}(2009){Atwood}, {Abdo}, {Ackermann}, {Althouse},
  {Anderson}, {Axelsson}, {Baldini}, {Ballet}, {Band}, {Barbiellini}, \&
  et~al.}]{2009ApJ...697.1071A}
{Atwood}, W.~B., {Abdo}, A.~A., {Ackermann}, M., {et~al.} 2009, ApJ, 697, 1071

\bibitem[{{Band} {et~al.}(1993){Band}, {Matteson}, {Ford}, {Schaefer},
  {Palmer}, {Teegarden}, {Cline}, {Briggs}, {Paciesas}, {Pendleton}, {Fishman},
  {Kouveliotou}, {Meegan}, {Wilson}, \& {Lestrade}}]{1993ApJ...413..281B}
{Band}, D., {Matteson}, J., {Ford}, L., {et~al.} 1993, ApJ, 413, 281

\bibitem[{{Bartolini} {et~al.}(2009){Bartolini}, {Greco}, {Guarnieri},
  {Piccioni}, {Beskin}, {Bondar}, {Karpov}, \&
  {Molinari}}]{2009arXiv0906.4144B}
{Bartolini}, C., {Greco}, G., {Guarnieri}, A., {et~al.} 2009, arXiv:0906.4144

\bibitem[{{Bernardini} {et~al.}(2007){Bernardini}, {Bianco}, {Caito},
  {Dainotti}, {Guida}, \& {Ruffini}}]{2007A&A...474L..13B}
{Bernardini}, M.~G., {Bianco}, C.~L., {Caito}, L., {et~al.} 2007, A\&A, 474,
  L13

\bibitem[{{Bernardini} {et~al.}(2008){Bernardini}, {Bianco}, {Caito},
  {Dainotti}, {Guida}, \& {Ruffini}}]{2008AIPC.1065..227B}
{Bernardini}, M.~G., {Bianco}, C.~L., {Caito}, L., {et~al.} 2008, in American
  Institute of Physics Conference Series, Vol. 1065, 2008 Nanjing Gamma-Ray
  Burst Conference, ed. Y.-F. {Huang}, Z.-G. {Dai}, \& B.~{Zhang}, 227--230

\bibitem[{{Bernardini} {et~al.}(2005){Bernardini}, {Bianco}, {Chardonnet},
  {Fraschetti}, {Ruffini}, \& {Xue}}]{2005ApJ...634L..29B}
{Bernardini}, M.~G., {Bianco}, C.~L., {Chardonnet}, P., {et~al.} 2005, ApJ,
  634, L29

\bibitem[{{Bernardini} {et~al.}(2009){Bernardini}, {Dainotti}, {Bianco},
  {Caito}, {Guida}, \& {Ruffini}}]{2009AIPC.1111..383B}
{Bernardini}, M.~G., {Dainotti}, M.~G., {Bianco}, C.~L., {et~al.} 2009, in
  American Institute of Physics Conference Series, Vol. 1111, Probing Stellar
  Populations out to the Distant Universe, ed. {G.~Giobbi, A.~Tornambe,
  G.~Raimondo, M.~Limongi, L.~A.~Antonelli, N.~Menci, \& E.~Brocato}, 383--386

\bibitem[{{Bianco} {et~al.}(2006){Bianco}, {Caito}, \&
  {Ruffini}}]{2006NCimB.121.1441B}
{Bianco}, C.~L., {Caito}, L., \& {Ruffini}, R. 2006, Nuovo Cimento B, 121, 1441

\bibitem[{{Bianco} \& {Ruffini}(2004)}]{2004ApJ...605L...1B}
{Bianco}, C.~L., \& {Ruffini}, R. 2004, ApJ, 605, L1

\bibitem[{{Bianco} \& {Ruffini}(2005{\natexlab{a}})}]{2005ApJ...633L..13B}
---. 2005{\natexlab{a}}, ApJ, 633, L13

\bibitem[{{Bianco} \& {Ruffini}(2005{\natexlab{b}})}]{2005ApJ...620L..23B}
---. 2005{\natexlab{b}}, ApJ, 620, L23

\bibitem[{{Blandford} \& {McKee}(1976)}]{1976PhFl...19.1130B}
{Blandford}, R.~D., \& {McKee}, C.~F. 1976, Physics of Fluids, 19, 1130

\bibitem[{{Bloom} {et~al.}(2006){Bloom}, {Perley}, \&
  {Chen}}]{2006GCN..5826....1B}
{Bloom}, J.~S., {Perley}, D.~A., \& {Chen}, H.~W. 2006, GCN Circ., 5826

\bibitem[{{Bo{\"e}r} {et~al.}(2006){Bo{\"e}r}, {Atteia}, {Damerdji}, {Gendre},
  {Klotz}, \& {Stratta}}]{2006ApJ...638L..71B}
{Bo{\"e}r}, M., {Atteia}, J.~L., {Damerdji}, Y., {et~al.} 2006, ApJ, 638, L71

\bibitem[{{Caito} {et~al.}(2010){Caito}, {Amati}, {Bernardini}, {Bianco}, {de
  Barros}, {Izzo}, {Patricelli}, \& {Ruffini}}]{2010A&A...521A..80C}
{Caito}, L., {Amati}, L., {Bernardini}, M.~G., {et~al.} 2010, A\&A, 521, A80

\bibitem[{{Caito} {et~al.}(2009){Caito}, {Bernardini}, {Bianco}, {Dainotti},
  {Guida}, \& {Ruffini}}]{2009A&A...498..501C}
{Caito}, L., {Bernardini}, M.~G., {Bianco}, C.~L., {et~al.} 2009, A\&A, 498,
  501

\bibitem[{{Cavallo} \& {Rees}(1978)}]{1978MNRAS.183..359C}
{Cavallo}, G., \& {Rees}, M.~J. 1978, MNRAS, 183, 359

\bibitem[{{Crider} {et~al.}(1998){Crider}, {Liang}, \&
  {Preece}}]{1998AIPC..428..359C}
{Crider}, A., {Liang}, E.~P., \& {Preece}, R.~D. 1998, in American Institute of
  Physics Conference Series, Vol. 428, Gamma-Ray Bursts, 4th Hunstville
  Symposium, ed. C.~A. {Meegan}, R.~D. {Preece}, \& T.~M. {Koshut}, 359--363

\bibitem[{{Crider} {et~al.}(1997){Crider}, {Liang}, {Smith}, {Preece},
  {Briggs}, {Pendleton}, {Paciesas}, {Band}, \&
  {Matteson}}]{1997ApJ...479L..39C}
{Crider}, A., {Liang}, E.~P., {Smith}, I.~A., {et~al.} 1997, ApJ, 479, L39

\bibitem[{{Cummings} {et~al.}(2005){Cummings}, {Angelini}, {Barthelmy},
  {Cucchiara}, {Gehrels}, {Gronwall}, {Holland}, {Mangano}, {Marshall},
  {Pagani}, \& {Palmer}}]{2005GCN..3910....1C}
{Cummings}, J., {Angelini}, L., {Barthelmy}, S., {et~al.} 2005, GCN Circ.,
  3910, 1

\bibitem[{{Curran} {et~al.}(2008){Curran}, {van der Horst}, \&
  {Wijers}}]{2008MNRAS.386..859C}
{Curran}, P.~A., {van der Horst}, A.~J., \& {Wijers}, R.~A.~M.~J. 2008, MNRAS,
  386, 859

\bibitem[{{Cusumano} {et~al.}(2007){Cusumano}, {Mangano}, {Chincarini},
  {Panaitescu}, {Burrows}, {La Parola}, {Sakamoto}, {Campana}, {Mineo},
  {Tagliaferri}, {Angelini}, {Barthelmy}, {Beardmore}, {Boyd}, {Cominsky},
  {Gronwall}, {Fenimore}, {Gehrels}, {Giommi}, {Goad}, {Hurley}, {Immler},
  {Kennea}, {Mason}, {Marshal}, {M{\'e}sz{\'a}ros}, {Nousek}, {Osborne},
  {Palmer}, {Roming}, {Wells}, {White}, \& {Zhang}}]{2007A&A...462...73C}
{Cusumano}, G., {Mangano}, V., {Chincarini}, G., {et~al.} 2007, A\&A, 462, 73

\bibitem[{{Daigne} {et~al.}(2009){Daigne}, {Bosnjak}, \&
  {Dubus}}]{2009arXiv0912.3743D}
{Daigne}, F., {Bosnjak}, Z., \& {Dubus}, G. 2009, ArXiv:0912.3743

\bibitem[{{Daigne} \& {Mochkovitch}(2002)}]{2002MNRAS.336.1271D}
{Daigne}, F., \& {Mochkovitch}, R. 2002, MNRAS, 336, 1271

\bibitem[{{Dainotti} {et~al.}(2007){Dainotti}, {Bernardini}, {Bianco}, {Caito},
  {Guida}, \& {Ruffini}}]{2007A&A...471L..29D}
{Dainotti}, M.~G., {Bernardini}, M.~G., {Bianco}, C.~L., {et~al.} 2007, A\&A,
  471, L29

\bibitem[{{Damour} \& {Ruffini}(1975)}]{1975PhRvL..35..463D}
{Damour}, T., \& {Ruffini}, R. 1975, Physical Review Letters, 35, 463

\bibitem[{{de Barros} {et~al.}(2011){de Barros}, {Amati}, {Bernardini},
  {Bianco}, {Caito}, {Izzo}, {Patricelli}, \& {Ruffini}}]{2011A&A...529A.130D}
{de Barros}, G., {Amati}, L., {Bernardini}, M.~G., {et~al.} 2011, A\&A, 529,
  A130

\bibitem[{{Dermer}(2006)}]{2006NCimB.121.1331D}
{Dermer}, C.~D. 2006, Nuovo Cimento B, 121, 1331

\bibitem[{{Dermer}(2008)}]{2008ApJ...684..430D}
---. 2008, ApJ, 684, 430

\bibitem[{{Dermer} \& {Mitman}(1999)}]{1999ApJ...513L...5D}
{Dermer}, C.~D., \& {Mitman}, K.~E. 1999, ApJ, 513, L5

\bibitem[{{Eichler} \& {Levinson}(2000)}]{2000ApJ...529..146E}
{Eichler}, D., \& {Levinson}, A. 2000, ApJ, 529, 146

\bibitem[{{Fermi}(1949)}]{1949PhRv...75.1169F}
{Fermi}, E. 1949, Physical Review, 75, 1169

\bibitem[{{Fermi}(1954)}]{1954ApJ...119....1F}
---. 1954, ApJ, 119, 1

\bibitem[{{Feroci} \& {The LOFT Consortium}(2011)}]{2011arXiv1107.0436F}
{Feroci}, M., \& {The LOFT Consortium}. 2011, arXiv:1107.0436

\bibitem[{{Finkelstein} {et~al.}(2009){Finkelstein}, {Rhoads}, {Malhotra}, \&
  {Grogin}}]{2009ApJ...691..465F}
{Finkelstein}, S.~L., {Rhoads}, J.~E., {Malhotra}, S., \& {Grogin}, N. 2009,
  ApJ, 691, 465

\bibitem[{{Frederiks} \& {Pal'Shin}(2011)}]{2011GCN..12370...1F}
{Frederiks}, D., \& {Pal'Shin}, V. 2011, GCN Circ., 12370, 1

\bibitem[{{Frontera} {et~al.}(2000){Frontera}, {Amati}, {Costa}, {Muller},
  {Pian}, {Piro}, {Soffitta}, {Tavani}, {Castro-Tirado}, {Dal Fiume}, {Feroci},
  {Heise}, {Masetti}, {Nicastro}, {Orlandini}, {Palazzi}, \&
  {Sari}}]{2000ApJS..127...59F}
{Frontera}, F., {Amati}, L., {Costa}, E., {et~al.} 2000, ApJSS, 127, 59

\bibitem[{{Gehrels} {et~al.}(2004){Gehrels}, {Chincarini}, {Giommi}, {Mason},
  {Nousek}, {Wells}, {White}, {Barthelmy}, {Burrows}, {Cominsky},
  {et~al.}}]{2004ApJ...611.1005G}
{Gehrels}, N., {Chincarini}, G., {Giommi}, P., {et~al.} 2004, ApJ, 611, 1005

\bibitem[{{Gendre} {et~al.}(2007){Gendre}, {Galli}, {Corsi}, {Klotz}, {Piro},
  {Stratta}, {Bo{\"e}r}, \& {Damerdji}}]{2007A&A...462..565G}
{Gendre}, B., {Galli}, A., {Corsi}, A., {et~al.} 2007, A\&A, 462, 565

\bibitem[{{Ghirlanda} {et~al.}(2002){Ghirlanda}, {Celotti}, \&
  {Ghisellini}}]{2002A&A...393..409G}
{Ghirlanda}, G., {Celotti}, A., \& {Ghisellini}, G. 2002, A\&A, 393, 409

\bibitem[{{Ghirlanda} {et~al.}(2003){Ghirlanda}, {Celotti}, \&
  {Ghisellini}}]{2003A&A...406..879G}
---. 2003, A\&A, 406, 879

\bibitem[{{Ghisellini} \& {Celotti}(1999)}]{1999A&AS..138..527G}
{Ghisellini}, G., \& {Celotti}, A. 1999, A\&AS, 138, 527

\bibitem[{{Giannios}(2006)}]{2006A&A...457..763G}
{Giannios}, D. 2006, A\&A, 457, 763

\bibitem[{{Golenetskii} {et~al.}(2006){Golenetskii}, {Aptekar}, {Mazets},
  {Pal'Shin}, {Frederiks}, \& {Cline}}]{2006GCN..5837....1G}
{Golenetskii}, S., {Aptekar}, R., {Mazets}, E., {et~al.} 2006, GCN Circ., 5837

\bibitem[{{Golenetskii} {et~al.}(2008){Golenetskii}, {Aptekar}, {Mazets},
  {Pal'Shin}, {Frederiks}, \& {Cline}}]{2008GCN..7482....1G}
---. 2008, GCN Circ., 7482

\bibitem[{{Goodman}(1986)}]{1986ApJ...308L..47G}
{Goodman}, J. 1986, ApJ, 308, L47

\bibitem[{{Granot} {et~al.}(1999){Granot}, {Piran}, \&
  {Sari}}]{1999ApJ...513..679G}
{Granot}, J., {Piran}, T., \& {Sari}, R. 1999, ApJ, 513, 679

\bibitem[{{Gruzinov} \& {Waxman}(1999)}]{1999ApJ...511..852G}
{Gruzinov}, A., \& {Waxman}, E. 1999, ApJ, 511, 852

\bibitem[{{Guidorzi} {et~al.}(2011){Guidorzi}, {Lacapra}, {Frontera},
  {Montanari}, {Amati}, {Calura}, {Nicastro}, \&
  {Orlandini}}]{2011A&A...526A..49G}
{Guidorzi}, C., {Lacapra}, M., {Frontera}, F., {et~al.} 2011, A\&A, 526, A49

\bibitem[{{Izzo} {et~al.}(2010){Izzo}, {Bernardini}, {Bianco}, {Caito},
  {Patricelli}, \& {Ruffini}}]{2010JKPS...57..551I}
{Izzo}, L., {Bernardini}, M.~G., {Bianco}, C.~L., {et~al.} 2010, Journal of
  Korean Physical Society, 57, 551

\bibitem[{{Izzo} {et~al.}(2012){Izzo}, {Ruffini}, {Penacchioni}, {Bianco},
  {Caito}, {Chakrabarti}, {Rueda}, {Nandi}, \&
  {Patricelli}}]{2012arXiv1202.4374I}
{Izzo}, L., {Ruffini}, R., {Penacchioni}, A.~V., {et~al.} 2012, ArXiv:1202.4374

\bibitem[{{Kaneko} {et~al.}(2008){Kaneko}, {Gonz{\'a}lez}, {Preece}, {Dingus},
  \& {Briggs}}]{2008ApJ...677.1168K}
{Kaneko}, Y., {Gonz{\'a}lez}, M.~M., {Preece}, R.~D., {Dingus}, B.~L., \&
  {Briggs}, M.~S. 2008, ApJ, 677, 1168

\bibitem[{{Kaneko} {et~al.}(2006){Kaneko}, {Preece}, {Briggs}, {Paciesas},
  {Meegan}, \& {Band}}]{2006ApJS..166..298K}
{Kaneko}, Y., {Preece}, R.~D., {Briggs}, M.~S., {et~al.} 2006, ApJSS, 166, 298

\bibitem[{{Kann} {et~al.}(2007){Kann}, {Masetti}, \&
  {Klose}}]{2007AJ....133.1187K}
{Kann}, D.~A., {Masetti}, N., \& {Klose}, S. 2007, AJ, 133, 1187

\bibitem[{{Kann} {et~al.}(2010){Kann}, {Klose}, {Zhang}, {Malesani}, {Nakar},
  {Pozanenko}, {Wilson}, {Butler}, {Jakobsson}, {Schulze}, {Andreev},
  {Antonelli}, {Bikmaev}, {Biryukov}, {B{\"o}ttcher}, {Burenin}, {Castro
  Cer{\'o}n}, {Castro-Tirado}, {Chincarini}, {Cobb}, {Covino}, {D'Avanzo},
  {D'Elia}, {Della Valle}, {de Ugarte Postigo}, {Efimov}, {Ferrero}, {Fugazza},
  {Fynbo}, {G{\aa}lfalk}, {Grundahl}, {Gorosabel}, {Gupta}, {Guziy}, {Hafizov},
  {Hjorth}, {Holhjem}, {Ibrahimov}, {Im}, {Israel}, {Je{\'l}inek}, {Jensen},
  {Karimov}, {Khamitov}, {Kizilo{\v g}lu}, {Klunko}, {Kub{\'a}nek}, {Kutyrev},
  {Laursen}, {Levan}, {Mannucci}, {Martin}, {Mescheryakov}, {Mirabal},
  {Norris}, {Ovaldsen}, {Paraficz}, {Pavlenko}, {Piranomonte}, {Rossi},
  {Rumyantsev}, {Salinas}, {Sergeev}, {Sharapov}, {Sollerman}, {Stecklum},
  {Stella}, {Tagliaferri}, {Tanvir}, {Telting}, {Testa}, {Updike}, {Volnova},
  {Watson}, {Wiersema}, \& {Xu}}]{2010ApJ...720.1513K}
{Kann}, D.~A., {Klose}, S., {Zhang}, B., {et~al.} 2010, ApJ, 720, 1513

\bibitem[{{Kawai} {et~al.}(2005){Kawai}, {Yamada}, {Kosugi}, {Hattori}, \&
  {Aoki}}]{2005GCN..3937....1K}
{Kawai}, N., {Yamada}, T., {Kosugi}, G., {Hattori}, T., \& {Aoki}, K. 2005, GCN
  Circ., 3937

\bibitem[{{Kim} {et~al.}(1998){Kim}, {Staveley-Smith}, {Sault}, {Kesteven},
  {McConnell}, {Dopita}, \& {Bessell}}]{1998PASA...15..132K}
{Kim}, S., {Staveley-Smith}, L., {Sault}, R.~J., {et~al.} 1998, Pub. Astron.
  Soc. Austr., 15, 132

\bibitem[{{Kumar}(1999)}]{1999ApJ...523L.113K}
{Kumar}, P. 1999, ApJ, 523, L113

\bibitem[{{Kumar} \& {McMahon}(2008)}]{2008MNRAS.384...33K}
{Kumar}, P., \& {McMahon}, E. 2008, MNRAS, 384, 33

\bibitem[{{Kumar} \& {Narayan}(2009)}]{2009MNRAS.395..472K}
{Kumar}, P., \& {Narayan}, R. 2009, MNRAS, 395, 472

\bibitem[{{Kumar} \& {Panaitescu}(2008)}]{2008MNRAS.391L..19K}
{Kumar}, P., \& {Panaitescu}, A. 2008, MNRAS, 391, L19

\bibitem[{{Lazzati} \& {Begelman}(2010)}]{2010ApJ...725.1137L}
{Lazzati}, D., \& {Begelman}, M.~C. 2010, ApJ, 725, 1137

\bibitem[{{Lazzati} {et~al.}(2000){Lazzati}, {Ghisellini}, {Celotti}, \&
  {Rees}}]{2000ApJ...529L..17L}
{Lazzati}, D., {Ghisellini}, G., {Celotti}, A., \& {Rees}, M.~J. 2000, ApJ,
  529, L17

\bibitem[{{Ledoux} {et~al.}(2006){Ledoux}, {Vreeswijk}, {Smette}, {Jaunsen}, \&
  {Kaufer}}]{2006GCN..5237....1L}
{Ledoux}, C., {Vreeswijk}, P., {Smette}, A., {Jaunsen}, A., \& {Kaufer}, A.
  2006, GCN Circ., 5237

\bibitem[{{Lockman}(2002)}]{2002ApJ...580L..47L}
{Lockman}, F.~J. 2002, ApJ, 580, L47

\bibitem[{{Margutti} {et~al.}(2011{\natexlab{a}}){Margutti}, {Bernardini},
  {Barniol Duran}, {Guidorzi}, {Shen}, \& {Chincarini}}]{2011MNRAS.410.1064M}
{Margutti}, R., {Bernardini}, M.~G., {Barniol Duran}, R., {et~al.}
  2011{\natexlab{a}}, MNRAS, 410, 1064

\bibitem[{{Margutti} {et~al.}(2008){Margutti}, {Guidorzi}, {Chincarini},
  {Pasotti}, {Covino}, \& {Mao}}]{2008AIPC.1065..259M}
{Margutti}, R., {Guidorzi}, C., {Chincarini}, G., {et~al.} 2008, in American
  Institute of Physics Conference Series, Vol. 1065, 2008 Nanjing Gamma-Ray
  Burst Conference, ed. Y.-F. {Huang}, Z.-G. {Dai}, \& B.~{Zhang}, 259--262

\bibitem[{{Margutti} {et~al.}(2011{\natexlab{b}}){Margutti}, {Chincarini},
  {Granot}, {Guidorzi}, {Berger}, {Bernardini}, {Gehrels}, {Soderberg},
  {Stamatikos}, \& {Zaninoni}}]{2011MNRAS.417.2144M}
{Margutti}, R., {Chincarini}, G., {Granot}, J., {et~al.} 2011{\natexlab{b}},
  MNRAS, 417, 2144

\bibitem[{{Medvedev}(2000)}]{2000ApJ...540..704M}
{Medvedev}, M.~V. 2000, ApJ, 540, 704

\bibitem[{{Medvedev} \& {Loeb}(1999)}]{1999ApJ...526..697M}
{Medvedev}, M.~V., \& {Loeb}, A. 1999, ApJ, 526, 697

\bibitem[{{Medvedev} \& {Spitkovsky}(2009)}]{2009ApJ...700..956M}
{Medvedev}, M.~V., \& {Spitkovsky}, A. 2009, ApJ, 700, 956

\bibitem[{{Meegan} {et~al.}(2009){Meegan}, {Lichti}, {Bhat}, {Bissaldi},
  {Briggs}, {Connaughton}, {Diehl}, {Fishman}, {Greiner}, {Hoover}, {van der
  Horst}, {von Kienlin}, {Kippen}, {Kouveliotou}, {McBreen}, {Paciesas},
  {Preece}, {Steinle}, {Wallace}, {Wilson}, \&
  {Wilson-Hodge}}]{2009ApJ...702..791M}
{Meegan}, C., {Lichti}, G., {Bhat}, P.~N., {et~al.} 2009, ApJ, 702, 791

\bibitem[{{M{\'e}sz{\'a}ros}(2002)}]{2002ARA&A..40..137M}
{M{\'e}sz{\'a}ros}, P. 2002, ARAA, 40, 137

\bibitem[{{Meszaros} {et~al.}(1993){Meszaros}, {Laguna}, \&
  {Rees}}]{1993ApJ...415..181M}
{Meszaros}, P., {Laguna}, P., \& {Rees}, M.~J. 1993, ApJ, 415, 181

\bibitem[{{M{\'e}sz{\'a}ros} {et~al.}(2002){M{\'e}sz{\'a}ros}, {Ramirez-Ruiz},
  {Rees}, \& {Zhang}}]{2002ApJ...578..812M}
{M{\'e}sz{\'a}ros}, P., {Ramirez-Ruiz}, E., {Rees}, M.~J., \& {Zhang}, B. 2002,
  ApJ, 578, 812

\bibitem[{{M{\'e}sz{\'a}ros} \& {Rees}(2000)}]{2000ApJ...530..292M}
{M{\'e}sz{\'a}ros}, P., \& {Rees}, M.~J. 2000, ApJ, 530, 292

\bibitem[{{Molinari} {et~al.}(2007){Molinari}, {Vergani}, {Malesani}, {Covino},
  {D'Avanzo}, {Chincarini}, {Zerbi}, {Antonelli}, {Conconi}, {Testa}, {Tosti},
  {Vitali}, {D'Alessio}, {Malaspina}, {Nicastro}, {Palazzi}, {Guetta},
  {Campana}, {Goldoni}, {Masetti}, {Meurs}, {Monfardini}, {Norci}, {Pian},
  {Piranomonte}, {Rizzuto}, {Stefanon}, {Stella}, {Tagliaferri}, {Ward},
  {Ihle}, {Gonzalez}, {Pizarro}, {Sinclaire}, \&
  {Valenzuela}}]{2007A&A...469L..13M}
{Molinari}, E., {Vergani}, S.~D., {Malesani}, D., {et~al.} 2007, A\&A, 469, L13

\bibitem[{{Nakar} \& {Granot}(2007)}]{2007MNRAS.380.1744N}
{Nakar}, E., \& {Granot}, J. 2007, MNRAS, 380, 1744

\bibitem[{{Paczynski}(1986)}]{1986ApJ...308L..43P}
{Paczynski}, B. 1986, ApJ, 308, L43

\bibitem[{{Page} {et~al.}(2006){Page}, {Starling}, {Osborne}, {Troja}, \&
  {Morris}}]{2006GCN..5832....1P}
{Page}, K.~L., {Starling}, R.~L.~C., {Osborne}, J.~P., {Troja}, E., \&
  {Morris}, D. 2006, GCN Circ., 5832

\bibitem[{{Page} {et~al.}(2007){Page}, {Willingale}, {Osborne}, {Zhang},
  {Godet}, {Marshall}, {Melandri}, {Norris}, {O'Brien}, {Pal'shin}, {Rol},
  {Romano}, {Starling}, {Schady}, {Yost}, {Barthelmy}, {Beardmore}, {Cusumano},
  {Burrows}, {De Pasquale}, {Ehle}, {Evans}, {Gehrels}, {Goad}, {Golenetskii},
  {Guidorzi}, {Mundell}, {Page}, {Ricker}, {Sakamoto}, {Schaefer},
  {Stamatikos}, {Troja}, {Ulanov}, {Yuan}, \&
  {Ziaeepour}}]{2007ApJ...663.1125P}
{Page}, K.~L., {Willingale}, R., {Osborne}, J.~P., {et~al.} 2007, ApJ, 663,
  1125

\bibitem[{{Panaitescu} \& {Meszaros}(1998)}]{1998ApJ...493L..31P}
{Panaitescu}, A., \& {Meszaros}, P. 1998, ApJ, 493, L31

\bibitem[{{Panaitescu} \& {M{\'e}sz{\'a}ros}(2000)}]{2000ApJ...544L..17P}
{Panaitescu}, A., \& {M{\'e}sz{\'a}ros}, P. 2000, ApJ, 544, L17

\bibitem[{{Pe'er} \& {Ryde}(2011)}]{2011ApJ...732...49P}
{Pe'er}, A., \& {Ryde}, F. 2011, ApJ, 732, 49

\bibitem[{{Pe'er} \& {Zhang}(2006)}]{2006ApJ...653..454P}
{Pe'er}, A., \& {Zhang}, B. 2006, ApJ, 653, 454

\bibitem[{{Penacchioni} {et~al.}(2012){Penacchioni}, {Ruffini}, {Izzo},
  {Muccino}, {Bianco}, {Caito}, {Patricelli}, \& {Amati}}]{2012A&A...538A..58P}
{Penacchioni}, A.~V., {Ruffini}, R., {Izzo}, L., {et~al.} 2012, A\&A, 538, A58

\bibitem[{{Piran}(1999)}]{1999PhR...314..575P}
{Piran}, T. 1999, Phys. Rep., 314, 575

\bibitem[{{Piran}(2004)}]{2004RvMP...76.1143P}
---. 2004, Reviews of Modern Physics, 76, 1143

\bibitem[{{Piran} {et~al.}(2009){Piran}, {Sari}, \&
  {Zou}}]{2009MNRAS.393.1107P}
{Piran}, T., {Sari}, R., \& {Zou}, Y. 2009, MNRAS, 393, 1107

\bibitem[{{Preece} {et~al.}(2002){Preece}, {Briggs}, {Giblin}, {Mallozzi},
  {Pendleton}, {Paciesas}, \& {Band}}]{2002ApJ...581.1248P}
{Preece}, R.~D., {Briggs}, M.~S., {Giblin}, T.~W., {et~al.} 2002, ApJ, 581,
  1248

\bibitem[{{Preece} {et~al.}(2000){Preece}, {Briggs}, {Mallozzi}, {Pendleton},
  {Paciesas}, \& {Band}}]{2000ApJS..126...19P}
{Preece}, R.~D., {Briggs}, M.~S., {Mallozzi}, R.~S., {et~al.} 2000, ApJSS, 126,
  19

\bibitem[{{Preece} {et~al.}(1998){Preece}, {Pendleton}, {Briggs}, {Mallozzi},
  {Paciesas}, {Band}, {Matteson}, \& {Meegan}}]{1998ApJ...496..849P}
{Preece}, R.~D., {Pendleton}, G.~N., {Briggs}, M.~S., {et~al.} 1998, ApJ, 496,
  849

\bibitem[{{Prochaska} {et~al.}(2004){Prochaska}, {Bloom}, {Chen}, {Hurley},
  {Melbourne}, {Dressler}, {Graham}, {Osip}, \& {Vacca}}]{2004ApJ...611..200P}
{Prochaska}, J.~X., {Bloom}, J.~S., {Chen}, H.-W., {et~al.} 2004, ApJ, 611, 200

\bibitem[{{Racusin} {et~al.}(2008){Racusin}, {Karpov}, {Sokolowski}, {Granot},
  {Wu}, {Pal'Shin}, {Covino}, {van der Horst}, {Oates}, {Schady}, {Smith},
  {Cummings}, {Starling}, {Piotrowski}, {Zhang}, {Evans}, {Holland}, {Malek},
  {Page}, {Vetere}, {Margutti}, {Guidorzi}, {Kamble}, {Curran}, {Beardmore},
  {Kouveliotou}, {Mankiewicz}, {Melandri}, {O'Brien}, {Page}, {Piran},
  {Tanvir}, {Wrochna}, {Aptekar}, {Barthelmy}, {Bartolini}, {Beskin}, {Bondar},
  {Bremer}, {Campana}, {Castro-Tirado}, {Cucchiara}, {Cwiok}, {D'Avanzo},
  {D'Elia}, {Della Valle}, {de Ugarte Postigo}, {Dominik}, {Falcone}, {Fiore},
  {Fox}, {Frederiks}, {Fruchter}, {Fugazza}, {Garrett}, {Gehrels},
  {Golenetskii}, {Gomboc}, {Gorosabel}, {Greco}, {Guarnieri}, {Immler},
  {Jelinek}, {Kasprowicz}, {La Parola}, {Levan}, {Mangano}, {Mazets},
  {Molinari}, {Moretti}, {Nawrocki}, {Oleynik}, {Osborne}, {Pagani}, {Pandey},
  {Paragi}, {Perri}, {Piccioni}, {Ramirez-Ruiz}, {Roming}, {Steele}, {Strom},
  {Testa}, {Tosti}, {Ulanov}, {Wiersema}, {Wijers}, {Winters}, {Zarnecki},
  {Zerbi}, {M{\'e}sz{\'a}ros}, {Chincarini}, \&
  {Burrows}}]{2008Natur.455..183R}
{Racusin}, J.~L., {Karpov}, S.~V., {Sokolowski}, M., {et~al.} 2008, Nature,
  455, 183

\bibitem[{{Ramirez-Ruiz} \& {Fenimore}(2000)}]{2000ApJ...539..712R}
{Ramirez-Ruiz}, E., \& {Fenimore}, E.~E. 2000, ApJ, 539, 712

\bibitem[{{Rees} \& {Meszaros}(1992)}]{1992MNRAS.258P..41R}
{Rees}, M.~J., \& {Meszaros}, P. 1992, MNRAS, 258, 41P

\bibitem[{{Rees} \& {Meszaros}(1994)}]{1994ApJ...430L..93R}
---. 1994, ApJ, 430, L93

\bibitem[{{Rees} \& {Meszaros}(1998)}]{1998ApJ...496L...1R}
---. 1998, ApJ, 496, L1

\bibitem[{{Ruffini} {et~al.}(2006){Ruffini}, {Bernardini}, {Bianco},
  {Chardonnet}, {Fraschetti}, {Guida}, \& {Xue}}]{2006ApJ...645L.109R}
{Ruffini}, R., {Bernardini}, M.~G., {Bianco}, C.~L., {et~al.} 2006, ApJ, 645,
  L109

\bibitem[{{Ruffini} {et~al.}(2004){Ruffini}, {Bianco}, {Chardonnet},
  {Fraschetti}, {Gurzadyan}, \& {Xue}}]{2004IJMPD..13..843R}
{Ruffini}, R., {Bianco}, C.~L., {Chardonnet}, P., {et~al.} 2004, IJMPD, 13, 843

\bibitem[{{Ruffini} {et~al.}(2005){Ruffini}, {Bianco}, {Chardonnet},
  {Fraschetti}, {Gurzadyan}, \& {Xue}}]{2005IJMPD..14...97R}
---. 2005, IJMPD, 14, 97

\bibitem[{{Ruffini} {et~al.}(2003){Ruffini}, {Bianco}, {Chardonnet},
  {Fraschetti}, {Vitagliano}, \& {Xue}}]{2003AIPC..668...16R}
{Ruffini}, R., {Bianco}, C.~L., {Chardonnet}, P., {et~al.} 2003, in American
  Institute of Physics Conference Series, Vol. 668, Cosmology and Gravitation,
  ed. M.~{Novello} \& S.~E. {Perez Bergliaffa}, 16--107

\bibitem[{{Ruffini} {et~al.}(2001{\natexlab{a}}){Ruffini}, {Bianco},
  {Chardonnet}, {Fraschetti}, \& {Xue}}]{2001ApJ...555L.113R}
{Ruffini}, R., {Bianco}, C.~L., {Chardonnet}, P., {Fraschetti}, F., \& {Xue},
  S.-S. 2001{\natexlab{a}}, ApJ, 555, L113

\bibitem[{{Ruffini} {et~al.}(2002){Ruffini}, {Bianco}, {Chardonnet},
  {Fraschetti}, \& {Xue}}]{2002ApJ...581L..19R}
---. 2002, ApJ, 581, L19

\bibitem[{{Ruffini} {et~al.}(2001{\natexlab{b}}){Ruffini}, {Bianco},
  {Fraschetti}, \& {Xue}}]{2001NCimB.116...99R}
{Ruffini}, R., {Bianco}, C.~L., {Fraschetti}, P.~C.~F., \& {Xue}, S.-S.
  2001{\natexlab{b}}, Nuovo Cimento B Serie, 116, 99

\bibitem[{{Ruffini} {et~al.}(2010{\natexlab{a}}){Ruffini}, {Chakrabarti}, \&
  {Izzo}}]{COSPAR}
{Ruffini}, R., {Chakrabarti}, S.~K., \& {Izzo}, L. 2010{\natexlab{a}},
  Submitted to Adv. Sp. Res.

\bibitem[{{Ruffini} {et~al.}(2011{\natexlab{a}}){Ruffini}, {Izzo},
  {Penacchioni}, {Bianco}, {Caito}, {Chakrabarti}, \& {Nandi}}]{TEXAS}
{Ruffini}, R., {Izzo}, L., {Penacchioni}, A.~V., {et~al.} 2011{\natexlab{a}},
  PoS(Texas2010), 101

\bibitem[{{Ruffini} {et~al.}(1999){Ruffini}, {Salmonson}, {Wilson}, \&
  {Xue}}]{1999A&A...350..334R}
{Ruffini}, R., {Salmonson}, J.~D., {Wilson}, J.~R., \& {Xue}, S.-S. 1999, A\&A,
  350, 334

\bibitem[{{Ruffini} {et~al.}(2000){Ruffini}, {Salmonson}, {Wilson}, \&
  {Xue}}]{2000A&A...359..855R}
---. 2000, A\&A, 359, 855

\bibitem[{{Ruffini} {et~al.}(2011{\natexlab{b}}){Ruffini}, {Siutsou}, \&
  {Vereshchagin}}]{2011arXiv1110.0407R}
{Ruffini}, R., {Siutsou}, I.~A., \& {Vereshchagin}, G.~V. 2011{\natexlab{b}},
  arXiv:1110.0407

\bibitem[{{Ruffini} {et~al.}(2010{\natexlab{b}}){Ruffini}, {Vereshchagin}, \&
  {Xue}}]{2010PhR...487....1R}
{Ruffini}, R., {Vereshchagin}, G., \& {Xue}, S. 2010{\natexlab{b}}, Phys. Rep.,
  487, 1

\bibitem[{{Ruffini} {et~al.}(2008){Ruffini}, {Aksenov}, {Bernardini}, {Bianco},
  {Caito}, {Dainotti}, {de Barros}, {Guida}, {Vereshchagin}, \&
  {Xue}}]{2008AIPC.1065..219R}
{Ruffini}, R., {Aksenov}, A.~G., {Bernardini}, M.~G., {et~al.} 2008, in
  American Institute of Physics Conference Series, Vol. 1065, 2008 Nanjing
  Gamma-Ray Burst Conference, ed. Y.-F. {Huang}, Z.-G. {Dai}, \& B.~{Zhang},
  219--222

\bibitem[{{Ruffini} {et~al.}(2009){Ruffini}, {Aksenov}, {Bernardini}, {Bianco},
  {Caito}, {Chardonnet}, {Dainotti}, {de Barros}, {Guida}, {Izzo},
  {Patricelli}, {Lemos}, {Rotondo}, {Hernandez}, {Vereshchagin}, \&
  {Xue}}]{2009AIPC.1132..199R}
{Ruffini}, R., {Aksenov}, A.~G., {Bernardini}, M.~G., {et~al.} 2009, in
  American Institute of Physics Conference Series, Vol. 1132, XIII Brazilian
  School on Cosmology and Gravitation, ed. M.~{Novello} \&
  S.~{Perez~Bergliaffa}, 199--266

\bibitem[{{Ryde}(2004)}]{2004ApJ...614..827R}
{Ryde}, F. 2004, ApJ, 614, 827

\bibitem[{{Ryde}(2005)}]{2005ApJ...625L..95R}
---. 2005, ApJ, 625, L95

\bibitem[{{Ryde} \& {Pe'er}(2009)}]{2009ApJ...702.1211R}
{Ryde}, F., \& {Pe'er}, A. 2009, ApJ, 702, 1211

\bibitem[{{Ryde} {et~al.}(2010){Ryde}, {Axelsson}, {Zhang}, {McGlynn}, {Pe'er},
  {Lundman}, {Larsson}, {Battelino}, {Zhang}, {Bissaldi}, {Bregeon}, {Briggs},
  {Chiang}, {de Palma}, {Guiriec}, {Larsson}, {Longo}, {McBreen}, {Omodei},
  {Petrosian}, {Preece}, \& {van der Horst}}]{2010ApJ...709L.172R}
{Ryde}, F., {Axelsson}, M., {Zhang}, B.~B., {et~al.} 2010, ApJ, 709, L172

\bibitem[{{Ryde} {et~al.}(2011){Ryde}, {Pe'Er}, {Nymark}, {Axelsson},
  {Moretti}, {Lundman}, {Battelino}, {Bissaldi}, {Chiang}, {Jackson},
  {Larsson}, {Longo}, {McGlynn}, \& {Omodei}}]{2011MNRAS.tmp..935R}
{Ryde}, F., {Pe'Er}, A., {Nymark}, T., {et~al.} 2011, MNRAS, 935

\bibitem[{{Rykoff} {et~al.}(2009){Rykoff}, {Aharonian}, {Akerlof}, {Ashley},
  {Barthelmy}, {Flewelling}, {Gehrels}, {G{\"o}{\v g}{\"u}{\c s}}, {G{\"u}ver},
  {Kizilo{\v g}lu}, {Krimm}, {McKay}, {{\"O}zel}, {Phillips}, {Quimby},
  {Rowell}, {Rujopakarn}, {Schaefer}, {Smith}, {Vestrand}, {Wheeler}, {Wren},
  {Yuan}, \& {Yost}}]{2009ApJ...702..489R}
{Rykoff}, E.~S., {Aharonian}, F., {Akerlof}, C.~W., {et~al.} 2009, ApJ, 702,
  489

\bibitem[{{Sakamoto} {et~al.}(2005){Sakamoto}, {Barbier}, {Barthelmy},
  {Cummings}, {Hullinger}, {Fenimore}, {Gehrels}, {Krimm}, {Markwardt},
  {Palmer}, {Parsons}, {Sato}, \& {Tueller}}]{2005GCN..3938....1S}
{Sakamoto}, T., {Barbier}, L., {Barthelmy}, S., {et~al.} 2005, GCN Circ., 3938

\bibitem[{{Sari}(1997)}]{1997ApJ...489L..37S}
{Sari}, R. 1997, ApJ, 489, L37

\bibitem[{{Sari}(1998)}]{1998ApJ...494L..49S}
---. 1998, ApJ, 494, L49

\bibitem[{{Sari} \& {Piran}(1999)}]{1999ApJ...520..641S}
{Sari}, R., \& {Piran}, T. 1999, ApJ, 520, 641

\bibitem[{{Schaefer} {et~al.}(1998){Schaefer}, {Palmer}, {Dingus}, {Schneid},
  {Schoenfelder}, {Ryan}, {Winkler}, {Hanlon}, {Kippen}, \&
  {Connors}}]{1998ApJ...492..696S}
{Schaefer}, B.~E., {Palmer}, D., {Dingus}, B.~L., {et~al.} 1998, ApJ, 492, 696

\bibitem[{{Shemi}(1994)}]{1994MNRAS.269.1112S}
{Shemi}, A. 1994, MNRAS, 269, 1112

\bibitem[{{Sironi} \& {Spitkovsky}(2009)}]{2009ApJ...707L..92S}
{Sironi}, L., \& {Spitkovsky}, A. 2009, ApJ, 707, L92

\bibitem[{{Spitkovsky}(2008{\natexlab{a}})}]{2008ApJ...673L..39S}
{Spitkovsky}, A. 2008{\natexlab{a}}, ApJ, 673, L39

\bibitem[{{Spitkovsky}(2008{\natexlab{b}})}]{2008ApJ...682L...5S}
---. 2008{\natexlab{b}}, ApJ, 682, L5

\bibitem[{{Stamatikos} {et~al.}(2009){Stamatikos}, {Ukwatta}, {Sakamoto},
  {Dhuga}, {Toma}, {Pe'Er}, {M{\'e}sz{\'a}ros}, {Band}, {Norris}, {Barthelmy},
  \& {Gehrels}}]{2009AIPC.1133..356S}
{Stamatikos}, M., {Ukwatta}, T.~N., {Sakamoto}, T., {et~al.} 2009, in American
  Institute of Physics Conference Series, Vol. 1133, GAMMA-RAY BURST: Sixth
  Huntsville Symposium, ed. C.~{Meegan}, C.~{Kouveliotou}, \& N.~{Gehrels},
  356--361

\bibitem[{{Stern} \& {Poutanen}(2004)}]{2004MNRAS.352L..35S}
{Stern}, B.~E., \& {Poutanen}, J. 2004, MNRAS, 352, L35

\bibitem[{{Sugita} {et~al.}(2009){Sugita}, {Yamaoka}, {Ohno}, {Tashiro},
  {Nakagawa}, {Urata}, {Pal'Shin}, {Golenetskii}, {Sakamoto}, {Cummings},
  {Krimm}, {Stamatikos}, {Parsons}, {Barthelmy}, \&
  {Gehrels}}]{2009PASJ...61..521S}
{Sugita}, S., {Yamaoka}, K., {Ohno}, M., {et~al.} 2009, PASJ, 61, 521

\bibitem[{{Tagliaferri} {et~al.}(2005){Tagliaferri}, {Antonelli}, {Chincarini},
  {Fern{\'a}ndez-Soto}, {Malesani}, {Della Valle}, {D'Avanzo}, {Grazian},
  {Testa}, {Campana}, {Covino}, {Fiore}, {Stella}, {Castro-Tirado},
  {Gorosabel}, {Burrows}, {Capalbi}, {Cusumano}, {Conciatore}, {D'Elia},
  {Filliatre}, {Fugazza}, {Gehrels}, {Goldoni}, {Guetta}, {Guziy}, {Held},
  {Hurley}, {Israel}, {Jel{\'{\i}}nek}, {Lazzati}, {L{\'o}pez-Echarri},
  {Melandri}, {Mirabel}, {Moles}, {Moretti}, {Mason}, {Nousek}, {Osborne},
  {Pellizza}, {Perna}, {Piranomonte}, {Piro}, {de Ugarte Postigo}, \&
  {Romano}}]{2005A&A...443L...1T}
{Tagliaferri}, G., {Antonelli}, L.~A., {Chincarini}, G., {et~al.} 2005, A\&A,
  443, L1

\bibitem[{{Tanvir} {et~al.}(2010){Tanvir}, {Rol}, {Levan}, {Svensson},
  {Fruchter}, {Granot}, {O'Brien}, {Wiersema}, {Starling}, {Jakobsson},
  {Fynbo}, {Hjorth}, {Curran}, {van der Horst}, {Kouveliotou}, {Racusin},
  {Burrows}, \& {Genet}}]{2010ApJ...725..625T}
{Tanvir}, N.~R., {Rol}, E., {Levan}, A.~J., {et~al.} 2010, ApJ, 725, 625

\bibitem[{{Tavani}(1996)}]{1996ApJ...466..768T}
{Tavani}, M. 1996, ApJ, 466, 768

\bibitem[{{Tavani} {et~al.}(2009){Tavani}, {Barbiellini}, {Argan}, {Boffelli},
  {Bulgarelli}, {Caraveo}, {Cattaneo}, {Chen}, {Cocco}, {Costa}, {D'Ammando},
  {Del Monte}, {de Paris}, {Di Cocco}, {di Persio}, {Donnarumma},
  {Evangelista}, {Feroci}, {Ferrari}, {Fiorini}, {Fornari}, {Fuschino},
  {Froysland}, {Frutti}, {Galli}, {Gianotti}, {Giuliani}, {Labanti}, {Lapshov},
  {Lazzarotto}, {Liello}, {Lipari}, {Longo}, {Mattaini}, {Marisaldi},
  {Mastropietro}, {Mauri}, {Mauri}, {Mereghetti}, {Morelli}, {Morselli},
  {Pacciani}, {Pellizzoni}, {Perotti}, {Piano}, {Picozza}, {Pontoni},
  {Porrovecchio}, {Prest}, {Pucella}, {Rapisarda}, {Rappoldi}, {Rossi},
  {Rubini}, {Soffitta}, {Traci}, {Trifoglio}, {Trois}, {Vallazza},
  {Vercellone}, {Vittorini}, {Zambra}, {Zanello}, {Pittori}, {Preger},
  {Santolamazza}, {Verrecchia}, {Giommi}, {Colafrancesco}, {Antonelli},
  {Cutini}, {Gasparrini}, {Stellato}, {Fanari}, {Primavera}, {Tamburelli},
  {Viola}, {Guarrera}, {Salotti}, {D'Amico}, {Marchetti}, {Crisconio},
  {Sabatini}, {Annoni}, {Alia}, {Longoni}, {Sanquerin}, {Battilana}, {Concari},
  {Dessimone}, {Grossi}, {Parise}, {Monzani}, {Artina}, {Pavesi},
  {Marseguerra}, {Nicolini}, {Scandelli}, {Soli}, {Vettorello}, {Zardetto},
  {Bonati}, {Maltecca}, {D'Alba}, {Patan{\'e}}, {Babini}, {Onorati},
  {Acquaroli}, {Angelucci}, {Morelli}, {Agostara}, {Cerone}, {Michetti},
  {Tempesta}, {D'Eramo}, {Rocca}, {Giannini}, {Borghi}, {Garavelli}, {Conte},
  {Balasini}, {Ferrario}, {Vanotti}, {Collavo}, \&
  {Giacomazzo}}]{2009A&A...502..995T}
{Tavani}, M., {Barbiellini}, G., {Argan}, A., {et~al.} 2009, A\&A, 502, 995

\bibitem[{{Ulanov} {et~al.}(2005){Ulanov}, {Golenetskii}, {Frederiks},
  {Mazets}, {Kokomov}, \& {Palshin}}]{2005NCimC..28..351U}
{Ulanov}, M.~V., {Golenetskii}, S.~V., {Frederiks}, D.~D., {et~al.} 2005, Nuovo
  Cimento C, 28, 351

\bibitem[{{van Paradijs} {et~al.}(2000){van Paradijs}, {Kouveliotou}, \&
  {Wijers}}]{2000ARA&A..38..379V}
{van Paradijs}, J., {Kouveliotou}, C., \& {Wijers}, R.~A.~M.~J. 2000, ARAA, 38,
  379

\bibitem[{{Vreeswijk} {et~al.}(2008){Vreeswijk}, {Smette}, {Malesani}, {Fynbo},
  {Milvang-Jensen}, {Jakobsson}, {Jaunsen}, {Oslo}, \&
  {Ledoux}}]{2008GCN..7444....1V}
{Vreeswijk}, P.~M., {Smette}, A., {Malesani}, D., {et~al.} 2008, GCN Circ.,
  7444

\bibitem[{{Waxman}(1997)}]{1997ApJ...491L..19W}
{Waxman}, E. 1997, ApJ, 491, L19

\bibitem[{{Wo{\'z}niak} {et~al.}(2009){Wo{\'z}niak}, {Vestrand}, {Panaitescu},
  {Wren}, {Davis}, \& {White}}]{2009ApJ...691..495W}
{Wo{\'z}niak}, P.~R., {Vestrand}, W.~T., {Panaitescu}, A.~D., {et~al.} 2009,
  ApJ, 691, 495

\bibitem[{{Yamaoka} {et~al.}(2009){Yamaoka}, {Endo}, {Enoto}, {Fukazawa},
  {Hara}, {Hanabata}, {Hong}, {Kamae}, {Kira}, {Kodaka}, {Kokubun}, {Maeno},
  {Makishima}, {Miyawaki}, {Morigami}, {Murakami}, {Nakagawa}, {Nakazawa},
  {Ohmori}, {Ohno}, {Onda}, {Sato}, {Sonoda}, {Sugita}, {Suzuki}, {Suzuki},
  {Tajima}, {Takahashi}, {Takahashi}, {Tanaka}, {Tamagawa}, {Tashiro},
  {Terada}, {Uehara}, {Urata}, {Yamauchi}, {Yoshida}, {Hurley}, {Pal'Shin},
  {Sakamoto}, \& {Cummings}}]{2009PASJ...61S..35Y}
{Yamaoka}, K., {Endo}, A., {Enoto}, T., {et~al.} 2009, PASJ, 61, 35

\bibitem[{{Zdziarski} {et~al.}(1991){Zdziarski}, {Svensson}, \&
  {Paczynski}}]{1991ApJ...366..343Z}
{Zdziarski}, A.~A., {Svensson}, R., \& {Paczynski}, B. 1991, ApJ, 366, 343

\bibitem[{{Zhang} {et~al.}(2006){Zhang}, {Fan}, {Dyks}, {Kobayashi},
  {M{\'e}sz{\'a}ros}, {Burrows}, {Nousek}, \& {Gehrels}}]{2006ApJ...642..354Z}
{Zhang}, B., {Fan}, Y.~Z., {Dyks}, J., {et~al.} 2006, ApJ, 642, 354

\bibitem[{{Zou} {et~al.}(2011){Zou}, {Fan}, \& {Piran}}]{2011ApJ...726L...2Z}
{Zou}, Y.-C., {Fan}, Y.-Z., \& {Piran}, T. 2011, ApJ, 726, L2

\bibitem[{{Zou} \& {Piran}(2010)}]{2010MNRAS.402.1854Z}
{Zou}, Y.-C., \& {Piran}, T. 2010, MNRAS, 402, 1854

\end{thebibliography}
\end{document}